\documentclass[aps,prl,showpacs,twocolumn,superscriptaddress,10]{revtex4-1}
\pdfpageattr {/Group << /S /Transparency /I true /CS /DeviceRGB>>}
\usepackage{graphicx}
\usepackage{amsmath}
\usepackage{bm}

\usepackage{multirow}
\usepackage{booktabs}
\usepackage{hyperref}
\hypersetup{%
pdftitle={Adiabatic amplification of plasmons and demons in 2D systems}, %
pdfauthor={Zhiyuan Sun, D.N. Basov, and M.M. Fogler},%
pdfpagemode={UseNone},%
pdfstartview={FitH},%
breaklinks=true,%
citecolor=blue,%
colorlinks=true,%
linkcolor=blue,%
urlcolor=blue}

\usepackage[T1]{fontenc}
\usepackage{fourier}
\usepackage{baskervald}

\newcommand{\unit}[1]{\,\mathrm{#1}} 

\graphicspath{{Image/}}

\begin{document}

\title{Adiabatic amplification of plasmons and demons in 2D systems}

\author{Zhiyuan Sun}
\affiliation{Department of Physics, University of California San Diego, 9500 Gilman Drive, La Jolla, California 92093}

\author{D. N. Basov}
\affiliation{Department of Physics, University of California San Diego, 9500 Gilman Drive, La Jolla, California 92093}
\affiliation{Department of Physics, Columbia University,
538 West 120th Street, New York, New York 10027}

\author{M. M. Fogler}
\affiliation{Department of Physics, University of California San Diego, 9500 Gilman Drive, La Jolla, California 92093}

\date{\today}

\begin{abstract}

We theoretically investigate charged collective modes in
a two-dimensional conductor with hot electrons
where the instantaneous mode frequencies gradually increase or decrease with time.
We show that the loss compensation or even amplification of the modes may occur.
We apply our theory to two types of collective modes in graphene, the plasmons and the energy waves, which can be probed in optical pump-probe experiments.

\end{abstract}

\maketitle

\emph{Introduction}.
Plasmons in metals, semiconductors, and other solid-state systems have been a topic of intensive research for over half a century~\cite{Pines1956}.
Plasmonics has found a number of technological applications in chemical sensing, light manipulation, and information processing.
Photoexcitation by ultrashort laser pulses~\cite{MacDonald2008, Baida2011, Ni2016} is one of the methods to generate plasmons. 
When the pulsed excitation is of high enough power,
it can modify material properties of either the plasmonic medium or its electromagnetic environment,
which is the principle underlying the emerging field of \textit{active} plasmonics~\cite{MacDonald2008, Hess2012, Stockman2013spp, Seren2015}.
For example, photoexcitation-induced population inversion
may permit plasmon loss compensation or amplification~\cite{Stockman2013spp, Hess2012}.
More often, plasmon lifetime remains quite short, e.g., tens of femtoseconds (fs) in noble metals,
which is an obstacle to applications.
In experiments using ultrafast optical pulses,
the plasmon frequency changes with time as the system relaxes back to equilibrium.
However, because of high damping, 
it has been customary to treat plasmonic response of the system as quasi-stationary during the plasmon lifetime.

Recently, graphene has emerged as a new plasmonic medium distinguished by record-high tunability and confinement~\cite{Chen2012, Fei2012}.
Combating damping remains a challenge;
however, plasmon quality factors as high as $Q \sim 30$ have been
demonstrated~\cite{Woessner2015, Ni2016} for graphene encapsulated in hexagonal boron nitride.
A new scientific frontier in graphene plasmonics is nonlinear~\cite{Cox2014, Mikhailov2014} and nonequilibrium dynamics probed in ultrafast optical experiments~\cite{Wagner2014, Brida2013}. 
Plasmon amplification through stimulated emission~\cite{Rana2008,
Apalkov2014} has been proposed theoretically and
plasmon switching by optical pumping has been demonstrated experimentally~\cite{Ni2016}.

These encouraging developments motivate us to study the regime
where the plasmon lifetime is comparable or longer than the characteristic relaxation time in a material.
Although this regime may or may not be realizable in graphene,
we consider this as a theoretical possibility.
Previously, collective modes in media undergoing adiabatic evolution have been discussed in theoretical
astrophysics~\cite{Morton2009}, plasma physics~\cite{Schmit.2010},
and general relativity~\cite{Grishchuk.1975}.
In this Letter,
we apply similar ideas to solid-state materials,
which are better suited for controlled experiments.
Our key finding is that loss compensation or even amplification can be a \textit{natural} outcome of the transient plasmon dynamics.
Additionally, we show that the same concept applies
to the energy wave in graphene~\cite{Phan2013, Briskot2015},
which is a collective mode similar to acoustic plasmons (or ``demons''~\cite{Pines1956}) in metals and semiconductors~\cite{Ruvalds1981, Pinczuk1981, Padmanabhan2014}
and also to ``cosmic sound'' in the early universe~\cite{Sunyaev1970}.

\emph{Qualitative picture}.
To model a nonequilibrium system under intense photoexcitation we assume that its electron temperature $T$ is much larger than the lattice temperature $T_l$. Such a hot-electron state typically forms in metals and semiconductor a few tens
of fs after optical pumping.
This rapid thermalization (that is, relaxation of the electron distribution to the Fermi-Dirac form with the temperature $T$)
is due to strong interactions of electrons with each other and with optical phonons.
Subsequently, $T$ gradually decreases toward $T_l$ at a much slower ``cooling'' rate measured in picoseconds (ps), predominantly due to emission of acoustic phonons.
If the plasmon dispersion depends on $T$,
plasmons propagating
in this transient state would have a slowly changing frequency.
We will show that such an adiabatic change of the plasmon frequency could induce adiabatic amplification of the plasmon amplitude.

\begin{figure}[t]
\includegraphics[width=3.3in]{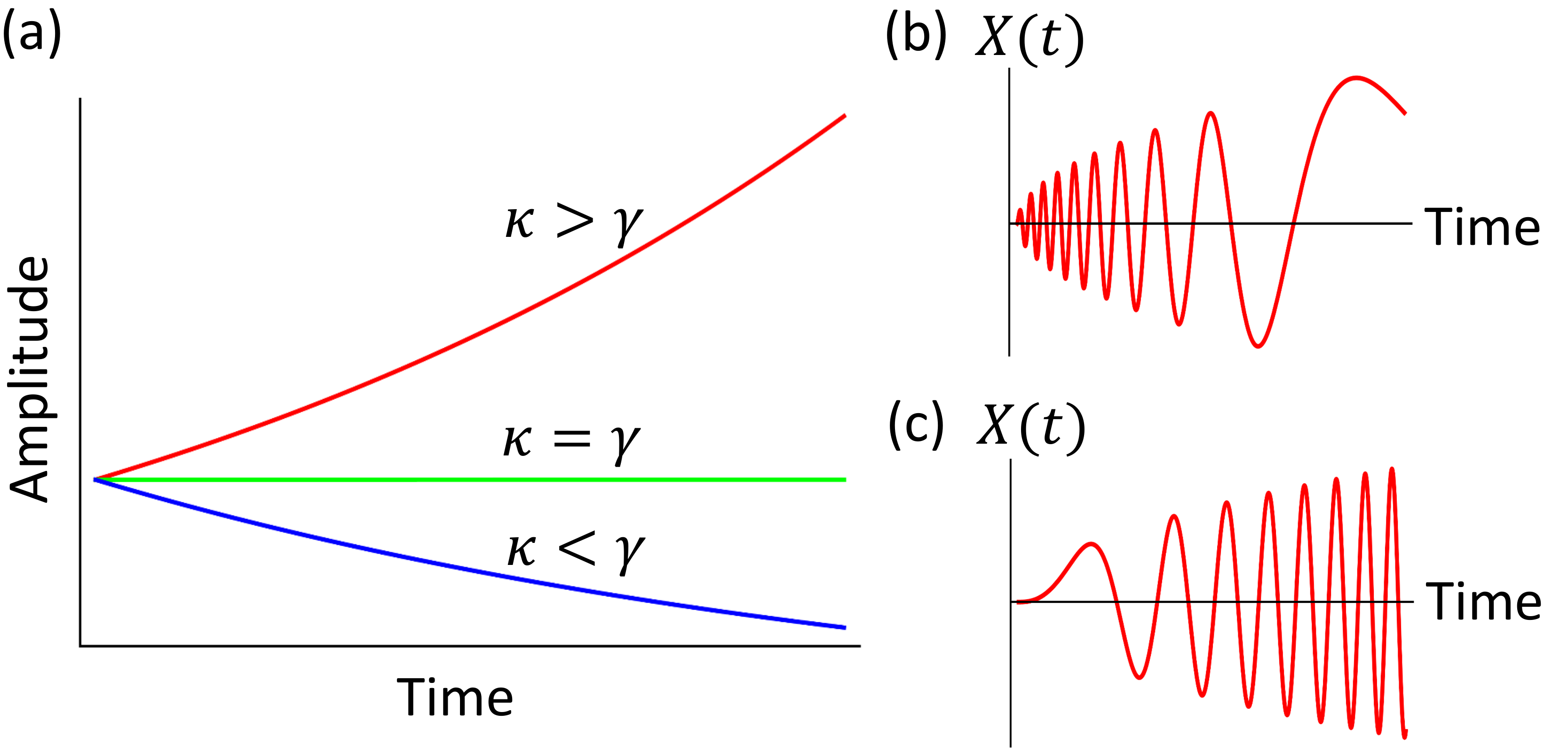} 
\caption{(a) A schematic showing the amplitude of a plasmon as a function of time $t$ for different relations between the mode frequency decay rate $\kappa$ and the damping rate $\gamma$. (b) The canonical coordinate $X$ as a function of $t$ in the $\kappa > \gamma$ case. The amplitude grows as the frequency drops.
(c) $X(t)$ for the case where amplification occurs while frequency increases, as in the tunneling process sketched in Fig.~\ref{fig:realization}(a) below.
}
\label{fig:amplitude_time_simplified}
\end{figure}

Adiabatic change of parameters has been previously considered in the context of plasmon-polariton focusing in tapered waveguides~\cite{Stockman2004}.
As plasmon approaches the narrow end of the waveguide,
its group velocity decreases and its electric field increases. 
In this situation the change of parameters occurs in space.
The mechanism we study relies instead on having parameters changing in time.
To explain our key idea let us treat the plasmon as a harmonic oscillator with the equation of motion
\begin{align}
\left( \partial^2_t  + \gamma(t) \partial_t  + \omega^2(t) \right) X = 0
\label{eqn:oscillator}
\end{align}
for its canonical coordinate $X(t)$ (e.g., charge density).
Here $\gamma(t)$ is the damping rate and $\omega(t)$ is the instantaneous mode frequency.
Suppose $\omega(t)$ changes monotonically with the decay rate $\kappa \equiv -\partial_t \ln \omega$, which is slow enough, $\omega \gg \kappa$,
then the Wentzel-Kramers-Brillouin (WKB) approximation to the solution of Eq.~\eqref{eqn:oscillator} is valid:
\begin{align}
X(t) &= A(t) e^{-i S(t)} \,,
\label{eqn:X}\\
A(t) &=  \frac{1}{\sqrt{\omega(t)}}
\exp\left(-\frac{1}{2} \int\limits_{0}^{t} \gamma(t_0) d t_0\right) \,,
\label{eqn:A}\\
S(t) &= \int\limits_{0}^{t} \omega(t_0) d t_0\,.
\label{eqn:S}
\end{align}
If both $\gamma$ and $\kappa$ are constant, the time-dependent plasmon amplitude has the form
\begin{align}
A(t) = e^{\frac{1}{2}(\kappa - \gamma)t } \,.
\label{eqn:amplitude_asymptotic2}
\end{align}
Clearly, the frequency decay rate $\kappa$ competes with the damping rate $\gamma$. If the condition
$
\kappa > \gamma
\label{eqn:amp_condition_II}
$
is met, then the oscillation amplitude increases with time, as shown in Figs.~\ref{fig:amplitude_time_simplified}(a) and (b).


Although the adiabatic principle appears simple and straightforward,
its application to actual solid-state systems may
require sorting out some important details. 
In the remainder of this Letter we do so
on the examples of two types of collective modes:
the plasmons and the energy waves in graphene.

\begin{figure}[b]
\includegraphics[width=2.5in]{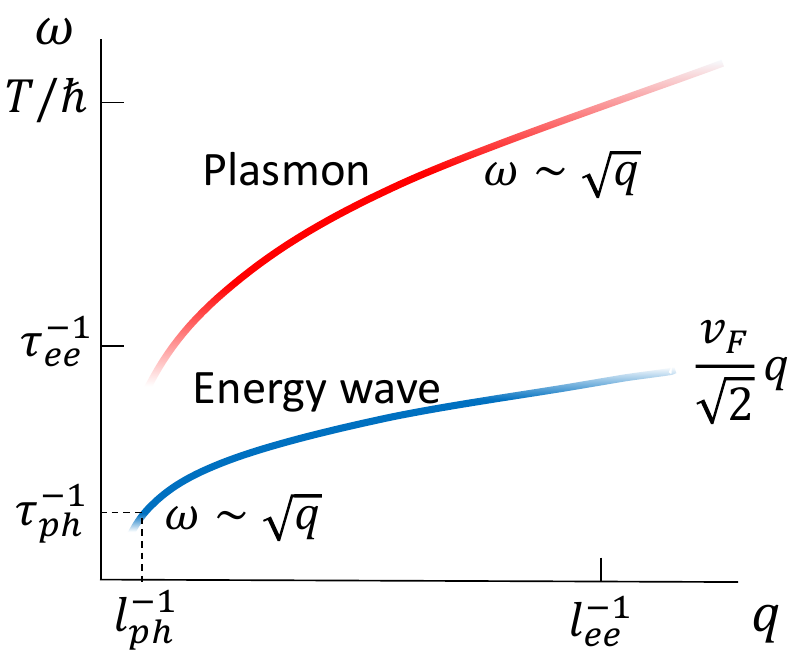} 
\caption{Dispersion of the plasmon~\cite{Vafek2006} and the energy wave~\cite{Phan2013, Briskot2015, Svintsov2012} in a weakly doped graphene with hot electrons (schematically). The plasmon (energy wave) exists at $\omega$ above (below) $\tau^{-1}_{ee}$; otherwise,
it is overdamped, as indicated by the fainting ends of the curves. 
For $T \gg \mu$ and $\alpha \ll 1$, $\tau^{-1}_{ee} = a \alpha^2 T$ with
$\alpha = {e^2} / {\epsilon \hbar v_F}$ and
$a \sim 4$~\cite{Kashuba2008, Muller2008a, Schuett2011, Briskot2015}.
The plasmon is also overdamped at $\omega \gtrsim T / \hbar$,
while the energy wave is damped by electron-phonon and disorder scattering characterized by the rate $\tau_{ph}^{-1}$.
}
\label{fig:parameter_region_hydro}
\end{figure}

\emph{Plasmons in two-dimensional (2D) materials}.
2D materials are very promising for active plasmonics because
they are not affected by a finite penetration length of optical beams and are much more tunable than bulk metals.
It is well known~\cite{Grigorenko2012, Basov2014, GarciadeAbajo2014} that such plasmons have a characteristic square-root dispersion with momentum
(Fig.~\ref{fig:parameter_region_hydro}),
$
\omega_q = \sqrt{\frac{2}{\epsilon}\, D q}\,,
\label{eqn:omega_2D}
$
where $\epsilon$ is the permittivity of the environment and $D$ is the Drude weight (see below).
Our goal is to show that the time-dependence of $D$ may give rise to adiabatic amplification of plasmons.

If the system has the spatial translational symmetry, different momenta are decoupled.
For a given $\mathbf{q}$,
in the linear-response regime, the plasmon dynamics is determined by the electrical conductivity operator with the kernel $\sigma_q(t, t_0)$.
Consider the following model for the conductivity kernel:
\begin{align}
\sigma_q(t, t_0) &= \frac{1}{\pi}\, D(t)
  e^{-\Gamma (t - t_0)} \theta(t - t_0) \,,
\label{eqn:sigma_q}\\
D(t) &= D(0) e^{-2 \int_{0}^{t} \kappa(t') d t'}.
\label{eqn:2kappa}
\end{align}
This model is motivated by a popular physical picture (see, for example, Refs.~\onlinecite{Orenstein2015}) where 
the current damping occurs because the ``density of photoexcited carriers'' decays with the rate $2\kappa$ and because of additionally,
these carriers experience momentum relaxation with the rate $\Gamma$,
see~\cite{SM} for further discussion.

Let us focus for now on the case of undoped graphene where
the Drude weight $D(t)$ is proportional to the electron temperature~\cite{Vafek2006}
$D(t) = 2 \ln 2\, \frac{e^2}{\hbar^2}\, T(t)$ and where
the underdamped plasmons exist at frequencies $\tau^{-1}_{ee} \ll \omega \ll T/\hbar$.
The lower limit is set by the electron-electron scattering rate $\tau^{-1}_{ee}$;
the upper limit is imposed by the Landau damping due to the interband transitions, see Fig.~\ref{fig:parameter_region_hydro}. 
In particular,
the dimensionless Landau damping rate of the thermal plasmons is given by~\cite{Schuett2011} $\Gamma / \omega = (\pi / 16 \ln 2)(\hbar\omega / T)^2$, which is small if $\hbar\omega \ll T$.
Note also that
the assumption of scalar $D$ can be justified if
$\kappa$ and $\gamma$ are much smaller than the electron-electron relaxation rate $\tau^{-1}_{ee}$ so that
an isotropic electron distribution (in the absence of a probe) is
maintained.

It is straightforward to show that the equation of motion for the plasmon has the same form as Eq.~\eqref{eqn:oscillator}
with $X$ equal to $\rho_q$,
the Fourier harmonic of the charge density,
and with the dissipation rate equal to
\begin{equation}
\gamma = 2\kappa + \Gamma\,.
\label{eqn:gamma}
\end{equation}
Unfortunately,
the condition $\kappa > \gamma$ seems impossible to satisfy since $\Gamma > 0$ and $\kappa > 0$.
In other words, the amplification cannot occur due to the plasmon damping rate being larger than the frequency decay rate, see also the supplementary material.

\begin{figure}[b]
\includegraphics[width=3.4 in]{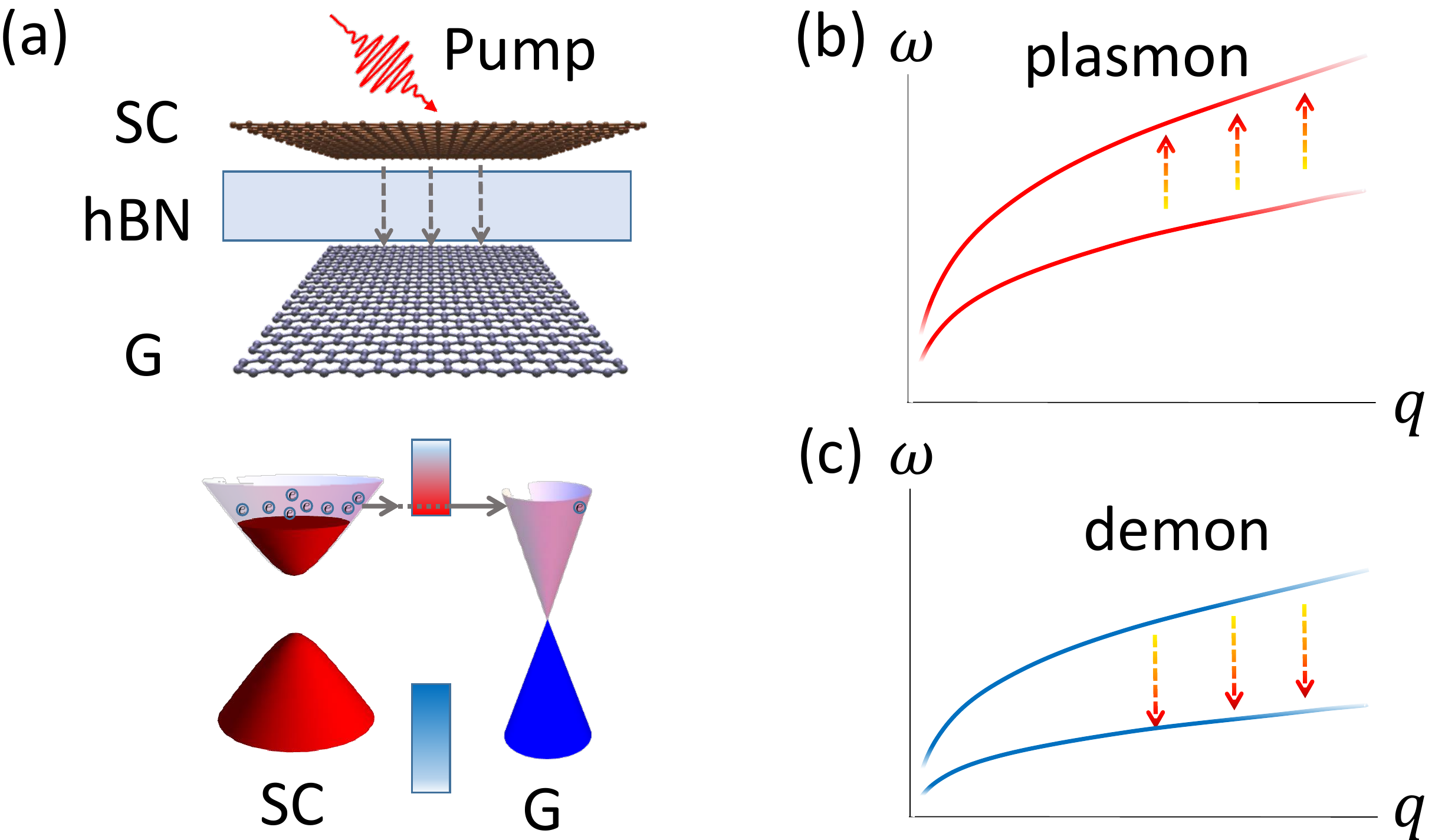} 
\caption{(a) Sketch of the heterostructure
made of parallel ultrathin layers of semiconductor (SC), insulator, and graphene (G)
where plasmon amplification can occur when hot electrons tunnel from the semiconductor to graphene. (b) A qualitative change of the plasmon dispersion during the tunneling. (c) A qualitative change of the demon dispersion with increasing temperature.
}
\label{fig:realization}
\end{figure}

Suppose, however, that the Drude weight is \textit{growing}, $\kappa < 0$.
In this case the criterion for amplification 
$
\kappa < -\Gamma
\label{eqn:amp_condition_I}
$
can be met if the growth rate is fast enough,
see Fig.~\ref{fig:parameter_region_hydro}(c).
Under what conditions can this scenario be realized?
One possibility is to leverage the dependence of the Drude weight on
the carrier density or effective mass,
which is another common attribute of ultrafast pump-probe experiments~\cite{MacDonald2008, Seren2015}.
We speculate that the plasmon amplification may be possible
by exploiting tunneling
in a vertical semiconductor/insulator/graphene heterostructure, see Fig.~\ref{fig:realization}(a).
The semiconductor could be, e.g., a transition-metal dichalcogenide
and the inslulator could be hexagonal boron nitride (hBN),
as in recent experiments~\cite{Massicotte2016}.
With a suitable bias voltage applied,
the initial state with a lower electrochemical potential in graphene can be maintained as the insulator bandgap would prevent electron tunneling in any direction. However, once they are heated to energies close or above the insulator's band edge,
the electrons in the semiconductor layer would tunnel to graphene.
(This is similar to a hot-electron doping effect~\cite{Fang2012pid} whereas in~\cite{Massicotte2016} the tunneling was in the opposite direction.)
For tunneling to be rapid the insulator must be thin,
which implies that the charges and current in the two layers
would also be coupled electromagnetically.
Therefore, the plasmons are the modes of the combined system.
If the effective carrier mass in the semiconductor is larger than that in graphene,
then the initial Drude weight is low but as a result of tunneling, the combined Drude weight of the carriers in the system (and hence, the electric current) would increase.
The upper limit for the amplification factor can be estimated
by completely neglecting the damping, $\Gamma \to 0$,
in the expression for the charge density amplitude
\begin{equation}
\rho_q(t) \propto e^{-\frac{1}{2}(\kappa + \Gamma)t },
\quad \kappa < 0\,.
\label{eqn:amplitude_3}
\end{equation}
The amplification is proportional to the square root of the plasmon frequency, or the fourth root of the Drude weight. If the increase of the latter comes from the decrease of the effective mass by, say, a factor of two, then plasmon amplification by as much as $\sim 20\%$ may be possible.
For more elaborate estimates, the carrier dynamics beyond the simple Drude approximation would need to be included in the model (see, for example, Ref.~\onlinecite{Huber2001} and the theory references cited therein).

The tunneling time of hot electrons across ultrathin hBN layers can as short as $7 \unit{fs}$ \cite{Ma2016},
which would correspond to $\kappa$ perhaps as high as several tens of $\mathrm{ps}^{-1}$.
In comparison,
the damping rate in hBN-encapsulated graphene was found to be $\Gamma \sim 2$ and $20 \unit{ps}^{-1}$ before and after the optical pump, respectively~\cite{Ni2016}. Hence, fulfilling the condition $\kappa > \gamma$ may be feasible.
Since the semiconductor would partially absorb the pump pulse,
graphene may remain relatively cool,
which may help reduce the plasmon damping due to electron-phonon scattering~\cite{Woessner2015}.

Experimental investigation of the
frequency, amplitude, and spatial
interference patterns of the amplified plasmons as a function of time may be possible by
far- and near-field pump-probe optical techniques~\cite{MacDonald2008, Ni2016,Fei2012}.

\emph{Energy wave (demon) in graphene}.
Our second example of a collective mode that may exhibit
adiabatic amplification is the energy wave in graphene.
This mode is predicted~\cite{Svintsov2012, Phan2013, Briskot2015} to exist in the hydrodynamic regime of
frequencies that are lower than the electron-electron collision rate
$\tau_{ee}^{-1}$, see Fig.~\ref{fig:parameter_region_hydro}.
In this regime, only collective variables immune to interparticle collisions,
i.e,
the zero modes of the collision integral are important:
the local temperature $T(\mathbf{r})$, chemical potential $\mu(\mathbf{r})$, and drift velocity $\mathbf{u}(\mathbf{r})$.
Their dynamics is described by a set of hydrodynamic equations~\cite{Muller2008a, Briskot2015, SM}. 
The energy wave
is the propagating longitudinal mode resulting from this set of equations.
Consider a weakly doped graphene, $\mu \ll T$.
The dispersion relation of the energy wave, neglecting dissipation, is
\begin{align}
\omega_q = v_F\sqrt{\frac{1}{2}\, q^2
 + \frac{4\pi}{3} \frac{e^2}{\epsilon}
 \frac{n^2}{n_E}\,  q} \,,
\label{eqn:energy_wave_disperison}
\end{align}
where $n$ is the average electron density and $n_E \equiv \langle\varepsilon\rangle $ is the average kinetic energy density (relative to the zero-doping, zero-temperature state).
The latter behaves as~\cite{SM}
$n_{E} \propto T^3$ in the regime we consider, $T \gg |\mu|$.
For $q \gg \frac{e^2}{\epsilon}
 \frac{n^2}{n_E}\equiv q_c$,
the dispersion of Eq.~\eqref{eqn:energy_wave_disperison} approaches $\omega=\frac{1}{\sqrt{2}}\, v_F q$.
This collective mode is neutral
because electrons and holes oscillate in phase.
It is similar to acoustic plasmons
observed in semiconductors~\cite{Pinczuk1981, Padmanabhan2014}.
Incidentally, the plasmons in Ref.~\cite{Alonso-Gonzalez2016}
were referred to as acoustic because their dispersion was
changed from the square-root to a linear one due to screening by a nearby gate.
This is unlike the original meaning of
the term acoustic plasmon (or ``demon'') introduced for 
a system where the screening is by electrons from a different band of the same material~\cite{Pines1956}.

For $q\ll q_c$,
the second term in the square root of Eq.~\eqref{eqn:energy_wave_disperison} dominates,
so $\omega_q \propto \sqrt{q}$. In this case the energy wave is no longer neutral: it involves both energy and charge density oscillations. However, it is different from the plasmon. First, the energy wave is in the hydrodynamic regime $\omega \ll \tau^{-1}_{ee}$ while the plasmon is in the high frequency regime $\omega \gg \tau^{-1}_{ee}$.
(In practice, the range of admissible $q$ is also limited from below by
the inverse mean-free path $l_{ph}$ due to electron-phonon and disorder scattering, see Fig.~\ref{fig:parameter_region_hydro}.)
Second, the frequency of the plasmon increases with electron temperature $T$ [similar to what is shown in Fig.~\ref{fig:realization}(b)] while that of the demons decreases [Fig.~\ref{fig:realization}(c)]. We will focus on the small $q$ region of the energy wave where its frequency can be efficiently controlled by $T$. For $\mu \sim 40\unit{meV} \approx 500\unit{K}$ and $T = 3000\unit{K}$, which is the regime probed in a recent experiment~\cite{Ni2016}, the wavelength corresponding to momentum $q_c$ is about $1 \,\mu \mathrm{m}$.
The change of temperature causes the change of the energy density $n_{E}$, which in turn affects the frequency of the energy wave through Eq.~\eqref{eqn:energy_wave_disperison}.
We assume that this change is adiabatic, in other words, that
the decay rate
$
\kappa = -\frac{1}{2}\, \partial_t \ln n_{E}\,
$
is a small parameter.
Keeping only the leading terms in the hydrodynamic equations,
we get the WKB solutions for the charge density $n_q$
and the energy density~\cite{SM}
\begin{align}
n_q(t) &\propto [n_{E}(t)]^{-1/4}\, e^{-i S(t)},
\label{eqn:charge_density_wkb}\\
n_{Eq}(t) &\propto [n_{E}(t)]^{3/4}\, e^{-i S(t)},
\label{eqn:energy_density_wkb}
\end{align}
with $S(t)$ given by Eq.~\eqref{eqn:S}.
Therefore, if we want to increase the energy density oscillations, we need to \textit{increase} the average energy density $n_E$; in other words, we need to heat up graphene.
This can be done using, for example, a moderate intensity laser source that heats the sample faster than the characteristic time $\tau_{ph}$ of
electron scattering by phonons and disorder. 
%
According to Eq. \eqref{eqn:energy_density_wkb},
the naive upper bound for the amplification factor (neglecting any damping) is $(T /T_l)^{9 / 4}$ where $T$ is the electron temperature after the photoexcitation. The amplification is only possible if graphene is slightly doped, in which case the energy mode is not purely neutral. Hence, it can also be probed by optical pump-probe spectroscopy, at THz frequencies. Alternatively,
it may be possible to exploit coupling of this mode to phonons and probe it by inelastic light scattering, similar to acoustic plasmons in semiconductors~\cite{Pinczuk1981, Padmanabhan2014}. 

\emph{Three-temperature state}.
A solid-state system exhibiting adiabatic amplification of collective modes would have another interesting nonequilibrium property. It
would have not two but three different effective temperatures. In addition to the lattice temperature $T_l$ and the electron temperature $T$,
it would also have the collective mode temperature $T_m$.
The temperature $T_m$ characterizes the modes created by random thermal fluctuations rather than those induced by an external probe pulse.
In the absence of damping, $\Gamma = 0$, the time evolution of $T_m$ can be deduced from the principle of entropy conservation in an adiabatic process.
The entropy of an ensemble of identical harmonic oscillators depends only on the ratio of temperature and their mode frequency.
The damping introduces an additional factor $e^{-\Gamma t}$.
Therefore, we expect
$
\frac{T_m(t)}{T_m(0)} \approx \frac{\omega_q(t)}{\omega_q(0)}\, e^{-\Gamma t}
\label{eqn:T_m}
$.
[Here we assume that $T_m$ is still much larger than the final equilibrium temperature $ T_m(t = \infty) = T_l$.]
When the hot-electron state is just created, $T_m = T_m(0)$ and $T = T(0)$ should presumably be of the same order.
Thereafter, they would diverge from one another.
For example, for graphene plasmons we find $T_m(t) \sim \omega_q(t) e^{-\Gamma t} \sim [T(t)]^{1/2} e^{-\Gamma t}$.
For graphene energy wave,
the same argument yields $T_m(t) \sim \omega_q(t) e^{-\Gamma t} \sim [T(t)]^{-3/2} e^{-\Gamma t}$.

In summary, we proposed the concept of adiabatic amplification of
collective modes in nonequilibrium systems under photoexcitation and
suggested two possible routes for its experimental realization in 2D materials.
Although we focused on systems with hot electrons,
the concept of adiabatic amplification is also applicable to
systems with ``cold'' electrons,
for example, superconducting films~\cite{Buisson1994, Stinson2014},
where plasmon damping can be even smaller than in graphene.

This work is supported by the U.S. Department of Energy under Grant DE-SC0012592.
We thank M.~K.~Liu and G.~Ni for discussions and also R.~D.~Averitt,
B.~N.~Narozhny, and M.~I.~Stockman for comments on the manuscript.

\nocite{Supplementary,Muller2008,Muller2009,Narozhny.2015}
\bibliography{Parametric_Amplification_MF}

\begin{thebibliography}{47}%
\makeatletter
\providecommand \@ifxundefined [1]{%
 \@ifx{#1\undefined}
}%
\providecommand \@ifnum [1]{%
 \ifnum #1\expandafter \@firstoftwo
 \else \expandafter \@secondoftwo
 \fi
}%
\providecommand \@ifx [1]{%
 \ifx #1\expandafter \@firstoftwo
 \else \expandafter \@secondoftwo
 \fi
}%
\providecommand \natexlab [1]{#1}%
\providecommand \enquote  [1]{``#1''}%
\providecommand \bibnamefont  [1]{#1}%
\providecommand \bibfnamefont [1]{#1}%
\providecommand \citenamefont [1]{#1}%
\providecommand \href@noop [0]{\@secondoftwo}%
\providecommand \href [0]{\begingroup \@sanitize@url \@href}%
\providecommand \@href[1]{\@@startlink{#1}\@@href}%
\providecommand \@@href[1]{\endgroup#1\@@endlink}%
\providecommand \@sanitize@url [0]{\catcode `\\12\catcode `\$12\catcode
  `\&12\catcode `\#12\catcode `\^12\catcode `\_12\catcode `\%12\relax}%
\providecommand \@@startlink[1]{}%
\providecommand \@@endlink[0]{}%
\providecommand \url  [0]{\begingroup\@sanitize@url \@url }%
\providecommand \@url [1]{\endgroup\@href {#1}{\urlprefix }}%
\providecommand \urlprefix  [0]{URL }%
\providecommand \Eprint [0]{\href }%
\providecommand \doibase [0]{http://dx.doi.org/}%
\providecommand \selectlanguage [0]{\@gobble}%
\providecommand \bibinfo  [0]{\@secondoftwo}%
\providecommand \bibfield  [0]{\@secondoftwo}%
\providecommand \translation [1]{[#1]}%
\providecommand \BibitemOpen [0]{}%
\providecommand \bibitemStop [0]{}%
\providecommand \bibitemNoStop [0]{.\EOS\space}%
\providecommand \EOS [0]{\spacefactor3000\relax}%
\providecommand \BibitemShut  [1]{\csname bibitem#1\endcsname}%
\let\auto@bib@innerbib\@empty
\bibitem [{\citenamefont {Pines}(1956)}]{Pines1956}%
  \BibitemOpen
  \bibfield  {author} {\bibinfo {author} {\bibfnamefont {D.}~\bibnamefont
  {Pines}},\ }\href {\doibase 10.1139/p56-154} {\bibfield  {journal} {\bibinfo
  {journal} {{C}an. {J}. {P}hys.}\ }\textbf {\bibinfo {volume} {34}},\ \bibinfo
  {pages} {1379} (\bibinfo {year} {1956})}\BibitemShut {NoStop}%
\bibitem [{\citenamefont {MacDonald}\ \emph {et~al.}(2008)\citenamefont
  {MacDonald}, \citenamefont {S\'{a}mson}, \citenamefont {Stockman},\ and\
  \citenamefont {Zheludev}}]{MacDonald2008}%
  \BibitemOpen
  \bibfield  {author} {\bibinfo {author} {\bibfnamefont {K.~F.}\ \bibnamefont
  {MacDonald}}, \bibinfo {author} {\bibfnamefont {Z.~L.}\ \bibnamefont
  {S\'{a}mson}}, \bibinfo {author} {\bibfnamefont {M.~I.}\ \bibnamefont
  {Stockman}}, \ and\ \bibinfo {author} {\bibfnamefont {N.~I.}\ \bibnamefont
  {Zheludev}},\ }\href {\doibase 10.1038/nphoton.2008.249} {\bibfield
  {journal} {\bibinfo  {journal} {{N}at. {P}hotonics}\ }\textbf {\bibinfo
  {volume} {3}},\ \bibinfo {pages} {55} (\bibinfo {year} {2008})}\BibitemShut
  {NoStop}%
\bibitem [{\citenamefont {Baida}\ \emph {et~al.}(2011)\citenamefont {Baida},
  \citenamefont {Mongin}, \citenamefont {Christofilos}, \citenamefont
  {Bachelier}, \citenamefont {Crut}, \citenamefont {Maioli}, \citenamefont
  {{Del Fatti}},\ and\ \citenamefont {Vall\'{e}e}}]{Baida2011}%
  \BibitemOpen
  \bibfield  {author} {\bibinfo {author} {\bibfnamefont {H.}~\bibnamefont
  {Baida}}, \bibinfo {author} {\bibfnamefont {D.}~\bibnamefont {Mongin}},
  \bibinfo {author} {\bibfnamefont {D.}~\bibnamefont {Christofilos}}, \bibinfo
  {author} {\bibfnamefont {G.}~\bibnamefont {Bachelier}}, \bibinfo {author}
  {\bibfnamefont {A.}~\bibnamefont {Crut}}, \bibinfo {author} {\bibfnamefont
  {P.}~\bibnamefont {Maioli}}, \bibinfo {author} {\bibfnamefont
  {N.}~\bibnamefont {{Del Fatti}}}, \ and\ \bibinfo {author} {\bibfnamefont
  {F.}~\bibnamefont {Vall\'{e}e}},\ }\href {\doibase
  10.1103/PhysRevLett.107.057402} {\bibfield  {journal} {\bibinfo  {journal}
  {{P}hys. {R}ev. {L}ett.}\ }\textbf {\bibinfo {volume} {107}},\ \bibinfo
  {pages} {057402} (\bibinfo {year} {2011})}\BibitemShut {NoStop}%
\bibitem [{\citenamefont {Ni}\ \emph {et~al.}(2016)\citenamefont {Ni},
  \citenamefont {Wang}, \citenamefont {Goldflam}, \citenamefont {Wagner},
  \citenamefont {Fei}, \citenamefont {McLeod}, \citenamefont {Liu},
  \citenamefont {Keilmann}, \citenamefont {{\"{O}}zyilmaz}, \citenamefont
  {{Castro Neto}}, \citenamefont {Hone}, \citenamefont {Fogler},\ and\
  \citenamefont {Basov}}]{Ni2016}%
  \BibitemOpen
  \bibfield  {author} {\bibinfo {author} {\bibfnamefont {G.~X.}\ \bibnamefont
  {Ni}}, \bibinfo {author} {\bibfnamefont {L.}~\bibnamefont {Wang}}, \bibinfo
  {author} {\bibfnamefont {M.~D.}\ \bibnamefont {Goldflam}}, \bibinfo {author}
  {\bibfnamefont {M.}~\bibnamefont {Wagner}}, \bibinfo {author} {\bibfnamefont
  {Z.}~\bibnamefont {Fei}}, \bibinfo {author} {\bibfnamefont {A.~S.}\
  \bibnamefont {McLeod}}, \bibinfo {author} {\bibfnamefont {M.~K.}\
  \bibnamefont {Liu}}, \bibinfo {author} {\bibfnamefont {F.}~\bibnamefont
  {Keilmann}}, \bibinfo {author} {\bibfnamefont {B.}~\bibnamefont
  {{\"{O}}zyilmaz}}, \bibinfo {author} {\bibfnamefont {A.~H.}\ \bibnamefont
  {{Castro Neto}}}, \bibinfo {author} {\bibfnamefont {J.}~\bibnamefont {Hone}},
  \bibinfo {author} {\bibfnamefont {M.~M.}\ \bibnamefont {Fogler}}, \ and\
  \bibinfo {author} {\bibfnamefont {D.~N.}\ \bibnamefont {Basov}},\ }\href
  {\doibase 10.1038/nphoton.2016.45} {\bibfield  {journal} {\bibinfo  {journal}
  {{N}at. {P}hotonics}\ }\textbf {\bibinfo {volume} {10}},\ \bibinfo {pages}
  {244} (\bibinfo {year} {2016})}\BibitemShut {NoStop}%
\bibitem [{\citenamefont {Hess}\ \emph {et~al.}(2012)\citenamefont {Hess},
  \citenamefont {Pendry}, \citenamefont {Maier}, \citenamefont {Oulton},
  \citenamefont {Hamm},\ and\ \citenamefont {Tsakmakidis}}]{Hess2012}%
  \BibitemOpen
  \bibfield  {author} {\bibinfo {author} {\bibfnamefont {O.}~\bibnamefont
  {Hess}}, \bibinfo {author} {\bibfnamefont {J.~B.}\ \bibnamefont {Pendry}},
  \bibinfo {author} {\bibfnamefont {S.~A.}\ \bibnamefont {Maier}}, \bibinfo
  {author} {\bibfnamefont {R.~F.}\ \bibnamefont {Oulton}}, \bibinfo {author}
  {\bibfnamefont {J.~M.}\ \bibnamefont {Hamm}}, \ and\ \bibinfo {author}
  {\bibfnamefont {K.~L.}\ \bibnamefont {Tsakmakidis}},\ }\href {\doibase
  10.1038/nmat3356} {\bibfield  {journal} {\bibinfo  {journal} {{N}at.
  {M}ater.}\ }\textbf {\bibinfo {volume} {11}},\ \bibinfo {pages} {573}
  (\bibinfo {year} {2012})}\BibitemShut {NoStop}%
\bibitem [{\citenamefont {Stockman}(2013)}]{Stockman2013spp}%
  \BibitemOpen
  \bibfield  {author} {\bibinfo {author} {\bibfnamefont {M.~I.}\ \bibnamefont
  {Stockman}},\ }in\ \href {\doibase 10.1002/9781118634394.ch1} {\emph
  {\bibinfo {booktitle} {Active Plasmonics And Tuneable Plasmonic
  Metamaterials}}},\ Vol.~\bibinfo {volume} {32},\ \bibinfo {editor} {edited
  by\ \bibinfo {editor} {\bibfnamefont {A.~V.}\ \bibnamefont {Zayats}}\ and\
  \bibinfo {editor} {\bibfnamefont {S.~A.}\ \bibnamefont {Maier}}}\ (\bibinfo
  {publisher} {Wiley},\ \bibinfo {year} {2013})\ pp.\ \bibinfo {pages}
  {1--39}\BibitemShut {NoStop}%
\bibitem [{\citenamefont {Seren}\ \emph {et~al.}(2016)\citenamefont {Seren},
  \citenamefont {Zhang}, \citenamefont {Keiser}, \citenamefont {Maddox},
  \citenamefont {Zhao}, \citenamefont {Fan}, \citenamefont {Bank},
  \citenamefont {Zhang},\ and\ \citenamefont {Averitt}}]{Seren2015}%
  \BibitemOpen
  \bibfield  {author} {\bibinfo {author} {\bibfnamefont {H.~R.}\ \bibnamefont
  {Seren}}, \bibinfo {author} {\bibfnamefont {J.}~\bibnamefont {Zhang}},
  \bibinfo {author} {\bibfnamefont {G.~R.}\ \bibnamefont {Keiser}}, \bibinfo
  {author} {\bibfnamefont {S.~J.}\ \bibnamefont {Maddox}}, \bibinfo {author}
  {\bibfnamefont {X.}~\bibnamefont {Zhao}}, \bibinfo {author} {\bibfnamefont
  {K.}~\bibnamefont {Fan}}, \bibinfo {author} {\bibfnamefont {S.~R.}\
  \bibnamefont {Bank}}, \bibinfo {author} {\bibfnamefont {X.}~\bibnamefont
  {Zhang}}, \ and\ \bibinfo {author} {\bibfnamefont {R.~D.}\ \bibnamefont
  {Averitt}},\ }\href {http://dx.doi.org/10.1038/lsa.2016.78} {\bibfield
  {journal} {\bibinfo  {journal} {Light Sci Appl.}\ }\textbf {\bibinfo {volume}
  {5}},\ \bibinfo {pages} {e16078} (\bibinfo {year} {2016})}\BibitemShut
  {NoStop}%
\bibitem [{\citenamefont {Chen}\ \emph {et~al.}(2012)\citenamefont {Chen},
  \citenamefont {Badioli}, \citenamefont {Alonso-Gonz\'{a}lez}, \citenamefont
  {Thongrattanasiri}, \citenamefont {Huth}, \citenamefont {Osmond},
  \citenamefont {Spasenovi\'{c}}, \citenamefont {Centeno}, \citenamefont
  {Pesquera}, \citenamefont {Godignon}, \citenamefont {Elorza}, \citenamefont
  {Camara}, \citenamefont {{Garc\'{\i}a de Abajo}}, \citenamefont
  {Hillenbrand},\ and\ \citenamefont {Koppens}}]{Chen2012}%
  \BibitemOpen
  \bibfield  {author} {\bibinfo {author} {\bibfnamefont {J.}~\bibnamefont
  {Chen}}, \bibinfo {author} {\bibfnamefont {M.}~\bibnamefont {Badioli}},
  \bibinfo {author} {\bibfnamefont {P.}~\bibnamefont {Alonso-Gonz\'{a}lez}},
  \bibinfo {author} {\bibfnamefont {S.}~\bibnamefont {Thongrattanasiri}},
  \bibinfo {author} {\bibfnamefont {F.}~\bibnamefont {Huth}}, \bibinfo {author}
  {\bibfnamefont {J.}~\bibnamefont {Osmond}}, \bibinfo {author} {\bibfnamefont
  {M.}~\bibnamefont {Spasenovi\'{c}}}, \bibinfo {author} {\bibfnamefont
  {A.}~\bibnamefont {Centeno}}, \bibinfo {author} {\bibfnamefont
  {A.}~\bibnamefont {Pesquera}}, \bibinfo {author} {\bibfnamefont
  {P.}~\bibnamefont {Godignon}}, \bibinfo {author} {\bibfnamefont {A.~Z.}\
  \bibnamefont {Elorza}}, \bibinfo {author} {\bibfnamefont {N.}~\bibnamefont
  {Camara}}, \bibinfo {author} {\bibfnamefont {F.~J.}\ \bibnamefont
  {{Garc\'{\i}a de Abajo}}}, \bibinfo {author} {\bibfnamefont {R.}~\bibnamefont
  {Hillenbrand}}, \ and\ \bibinfo {author} {\bibfnamefont {F.~H.~L.}\
  \bibnamefont {Koppens}},\ }\href {\doibase 10.1038/nature11254} {\bibfield
  {journal} {\bibinfo  {journal} {{N}ature}\ }\textbf {\bibinfo {volume}
  {487}},\ \bibinfo {pages} {77} (\bibinfo {year} {2012})}\BibitemShut
  {NoStop}%
\bibitem [{\citenamefont {Fei}\ \emph {et~al.}(2012)\citenamefont {Fei},
  \citenamefont {Rodin}, \citenamefont {Andreev}, \citenamefont {Bao},
  \citenamefont {McLeod}, \citenamefont {Wagner}, \citenamefont {Zhang},
  \citenamefont {Zhao}, \citenamefont {Thiemens}, \citenamefont {Dominguez},
  \citenamefont {Fogler}, \citenamefont {{Castro Neto}}, \citenamefont {Lau},
  \citenamefont {Keilmann},\ and\ \citenamefont {Basov}}]{Fei2012}%
  \BibitemOpen
  \bibfield  {author} {\bibinfo {author} {\bibfnamefont {Z.}~\bibnamefont
  {Fei}}, \bibinfo {author} {\bibfnamefont {A.~S.}\ \bibnamefont {Rodin}},
  \bibinfo {author} {\bibfnamefont {G.~O.}\ \bibnamefont {Andreev}}, \bibinfo
  {author} {\bibfnamefont {W.}~\bibnamefont {Bao}}, \bibinfo {author}
  {\bibfnamefont {A.~S.}\ \bibnamefont {McLeod}}, \bibinfo {author}
  {\bibfnamefont {M.}~\bibnamefont {Wagner}}, \bibinfo {author} {\bibfnamefont
  {L.~M.}\ \bibnamefont {Zhang}}, \bibinfo {author} {\bibfnamefont
  {Z.}~\bibnamefont {Zhao}}, \bibinfo {author} {\bibfnamefont {M.}~\bibnamefont
  {Thiemens}}, \bibinfo {author} {\bibfnamefont {G.}~\bibnamefont {Dominguez}},
  \bibinfo {author} {\bibfnamefont {M.~M.}\ \bibnamefont {Fogler}}, \bibinfo
  {author} {\bibfnamefont {A.~H.}\ \bibnamefont {{Castro Neto}}}, \bibinfo
  {author} {\bibfnamefont {C.~N.}\ \bibnamefont {Lau}}, \bibinfo {author}
  {\bibfnamefont {F.}~\bibnamefont {Keilmann}}, \ and\ \bibinfo {author}
  {\bibfnamefont {D.~N.}\ \bibnamefont {Basov}},\ }\href {\doibase
  10.1038/nature11253} {\bibfield  {journal} {\bibinfo  {journal} {{N}ature}\
  }\textbf {\bibinfo {volume} {487}},\ \bibinfo {pages} {82} (\bibinfo {year}
  {2012})}\BibitemShut {NoStop}%
\bibitem [{\citenamefont {Woessner}\ \emph {et~al.}(2015)\citenamefont
  {Woessner}, \citenamefont {Lundeberg}, \citenamefont {Gao}, \citenamefont
  {Principi}, \citenamefont {Alonso-Gonz\'{a}lez}, \citenamefont {Carrega},
  \citenamefont {Watanabe}, \citenamefont {Taniguchi}, \citenamefont {Vignale},
  \citenamefont {Polini}, \citenamefont {Hone}, \citenamefont {Hillenbrand},\
  and\ \citenamefont {Koppens}}]{Woessner2015}%
  \BibitemOpen
  \bibfield  {author} {\bibinfo {author} {\bibfnamefont {A.}~\bibnamefont
  {Woessner}}, \bibinfo {author} {\bibfnamefont {M.~B.}\ \bibnamefont
  {Lundeberg}}, \bibinfo {author} {\bibfnamefont {Y.}~\bibnamefont {Gao}},
  \bibinfo {author} {\bibfnamefont {A.}~\bibnamefont {Principi}}, \bibinfo
  {author} {\bibfnamefont {P.}~\bibnamefont {Alonso-Gonz\'{a}lez}}, \bibinfo
  {author} {\bibfnamefont {M.}~\bibnamefont {Carrega}}, \bibinfo {author}
  {\bibfnamefont {K.}~\bibnamefont {Watanabe}}, \bibinfo {author}
  {\bibfnamefont {T.}~\bibnamefont {Taniguchi}}, \bibinfo {author}
  {\bibfnamefont {G.}~\bibnamefont {Vignale}}, \bibinfo {author} {\bibfnamefont
  {M.}~\bibnamefont {Polini}}, \bibinfo {author} {\bibfnamefont
  {J.}~\bibnamefont {Hone}}, \bibinfo {author} {\bibfnamefont {R.}~\bibnamefont
  {Hillenbrand}}, \ and\ \bibinfo {author} {\bibfnamefont {F.~H.~L.}\
  \bibnamefont {Koppens}},\ }\href {\doibase 10.1038/nmat4169} {\bibfield
  {journal} {\bibinfo  {journal} {{N}at. {M}ater.}\ }\textbf {\bibinfo {volume}
  {14}},\ \bibinfo {pages} {421} (\bibinfo {year} {2015})}\BibitemShut
  {NoStop}%
\bibitem [{\citenamefont {Cox}\ and\ \citenamefont {{Javier Garc\'{\i}a de
  Abajo}}(2014)}]{Cox2014}%
  \BibitemOpen
  \bibfield  {author} {\bibinfo {author} {\bibfnamefont {J.~D.}\ \bibnamefont
  {Cox}}\ and\ \bibinfo {author} {\bibfnamefont {F.}~\bibnamefont {{Javier
  Garc\'{\i}a de Abajo}}},\ }\href {\doibase 10.1038/ncomms6725} {\bibfield
  {journal} {\bibinfo  {journal} {{N}at. {C}ommun.}\ }\textbf {\bibinfo
  {volume} {5}},\ \bibinfo {pages} {5725} (\bibinfo {year} {2014})}\BibitemShut
  {NoStop}%
\bibitem [{\citenamefont {Mikhailov}(2014)}]{Mikhailov2014}%
  \BibitemOpen
  \bibfield  {author} {\bibinfo {author} {\bibfnamefont {S.~A.}\ \bibnamefont
  {Mikhailov}},\ }\href {\doibase 10.1103/PhysRevB.90.241301} {\bibfield
  {journal} {\bibinfo  {journal} {{P}hys. {R}ev. {B}}\ }\textbf {\bibinfo
  {volume} {90}},\ \bibinfo {pages} {241301} (\bibinfo {year}
  {2014})}\BibitemShut {NoStop}%
\bibitem [{\citenamefont {Wagner}\ \emph {et~al.}(2014)\citenamefont {Wagner},
  \citenamefont {Fei}, \citenamefont {McLeod}, \citenamefont {Rodin},
  \citenamefont {Bao}, \citenamefont {Iwinski}, \citenamefont {Zhao},
  \citenamefont {Goldflam}, \citenamefont {Liu}, \citenamefont {Dominguez},
  \citenamefont {Thiemens}, \citenamefont {Fogler}, \citenamefont {{Castro
  Neto}}, \citenamefont {Lau}, \citenamefont {Amarie}, \citenamefont
  {Keilmann},\ and\ \citenamefont {Basov}}]{Wagner2014}%
  \BibitemOpen
  \bibfield  {author} {\bibinfo {author} {\bibfnamefont {M.}~\bibnamefont
  {Wagner}}, \bibinfo {author} {\bibfnamefont {Z.}~\bibnamefont {Fei}},
  \bibinfo {author} {\bibfnamefont {A.~S.}\ \bibnamefont {McLeod}}, \bibinfo
  {author} {\bibfnamefont {A.~S.}\ \bibnamefont {Rodin}}, \bibinfo {author}
  {\bibfnamefont {W.}~\bibnamefont {Bao}}, \bibinfo {author} {\bibfnamefont
  {E.~G.}\ \bibnamefont {Iwinski}}, \bibinfo {author} {\bibfnamefont
  {Z.}~\bibnamefont {Zhao}}, \bibinfo {author} {\bibfnamefont {M.}~\bibnamefont
  {Goldflam}}, \bibinfo {author} {\bibfnamefont {M.}~\bibnamefont {Liu}},
  \bibinfo {author} {\bibfnamefont {G.}~\bibnamefont {Dominguez}}, \bibinfo
  {author} {\bibfnamefont {M.}~\bibnamefont {Thiemens}}, \bibinfo {author}
  {\bibfnamefont {M.~M.}\ \bibnamefont {Fogler}}, \bibinfo {author}
  {\bibfnamefont {A.~H.}\ \bibnamefont {{Castro Neto}}}, \bibinfo {author}
  {\bibfnamefont {C.~N.}\ \bibnamefont {Lau}}, \bibinfo {author} {\bibfnamefont
  {S.}~\bibnamefont {Amarie}}, \bibinfo {author} {\bibfnamefont
  {F.}~\bibnamefont {Keilmann}}, \ and\ \bibinfo {author} {\bibfnamefont
  {D.~N.}\ \bibnamefont {Basov}},\ }\href {\doibase 10.1021/nl4042577}
  {\bibfield  {journal} {\bibinfo  {journal} {{N}ano {L}ett.}\ }\textbf
  {\bibinfo {volume} {14}},\ \bibinfo {pages} {894} (\bibinfo {year}
  {2014})}\BibitemShut {NoStop}%
\bibitem [{\citenamefont {Brida}\ \emph {et~al.}(2013)\citenamefont {Brida},
  \citenamefont {Tomadin}, \citenamefont {Manzoni}, \citenamefont {Kim},
  \citenamefont {Lombardo}, \citenamefont {Milana}, \citenamefont {Nair},
  \citenamefont {Novoselov}, \citenamefont {Ferrari}, \citenamefont {Cerullo},\
  and\ \citenamefont {Polini}}]{Brida2013}%
  \BibitemOpen
  \bibfield  {author} {\bibinfo {author} {\bibfnamefont {D.}~\bibnamefont
  {Brida}}, \bibinfo {author} {\bibfnamefont {A.}~\bibnamefont {Tomadin}},
  \bibinfo {author} {\bibfnamefont {C.}~\bibnamefont {Manzoni}}, \bibinfo
  {author} {\bibfnamefont {Y.~J.}\ \bibnamefont {Kim}}, \bibinfo {author}
  {\bibfnamefont {A.}~\bibnamefont {Lombardo}}, \bibinfo {author}
  {\bibfnamefont {S.}~\bibnamefont {Milana}}, \bibinfo {author} {\bibfnamefont
  {R.~R.}\ \bibnamefont {Nair}}, \bibinfo {author} {\bibfnamefont {K.~S.}\
  \bibnamefont {Novoselov}}, \bibinfo {author} {\bibfnamefont {A.~C.}\
  \bibnamefont {Ferrari}}, \bibinfo {author} {\bibfnamefont {G.}~\bibnamefont
  {Cerullo}}, \ and\ \bibinfo {author} {\bibfnamefont {M.}~\bibnamefont
  {Polini}},\ }\href {\doibase 10.1038/ncomms2987} {\bibfield  {journal}
  {\bibinfo  {journal} {{N}at. {C}ommun.}\ }\textbf {\bibinfo {volume} {4}},\
  \bibinfo {pages} {1987} (\bibinfo {year} {2013})}\BibitemShut {NoStop}%
\bibitem [{\citenamefont {Rana}(2008)}]{Rana2008}%
  \BibitemOpen
  \bibfield  {author} {\bibinfo {author} {\bibfnamefont {F.}~\bibnamefont
  {Rana}},\ }\href {\doibase 10.1109/TNANO.2007.910334} {\bibfield  {journal}
  {\bibinfo  {journal} {{IEEE} {T}rans. {N}anotechnol.}\ }\textbf {\bibinfo
  {volume} {7}},\ \bibinfo {pages} {91} (\bibinfo {year} {2008})}\BibitemShut
  {NoStop}%
\bibitem [{\citenamefont {Apalkov}\ and\ \citenamefont
  {Stockman}(2014)}]{Apalkov2014}%
  \BibitemOpen
  \bibfield  {author} {\bibinfo {author} {\bibfnamefont {V.}~\bibnamefont
  {Apalkov}}\ and\ \bibinfo {author} {\bibfnamefont {M.~I.}\ \bibnamefont
  {Stockman}},\ }\href {\doibase 10.1038/lsa.2014.72} {\bibfield  {journal}
  {\bibinfo  {journal} {{L}ight {S}ci. {A}ppl.}\ }\textbf {\bibinfo {volume}
  {3}},\ \bibinfo {pages} {e191} (\bibinfo {year} {2014})}\BibitemShut
  {NoStop}%
\bibitem [{\citenamefont {Morton}\ \emph {et~al.}(2010)\citenamefont {Morton},
  \citenamefont {Hood},\ and\ \citenamefont {Erd\'{e}lyi}}]{Morton2009}%
  \BibitemOpen
  \bibfield  {author} {\bibinfo {author} {\bibfnamefont {R.~J.}\ \bibnamefont
  {Morton}}, \bibinfo {author} {\bibfnamefont {A.~W.}\ \bibnamefont {Hood}}, \
  and\ \bibinfo {author} {\bibfnamefont {R.}~\bibnamefont {Erd\'{e}lyi}},\
  }\href {\doibase 10.1051/0004-6361/200913365} {\bibfield  {journal} {\bibinfo
   {journal} {{A}stron. {A}strophys.}\ }\textbf {\bibinfo {volume} {512}},\
  \bibinfo {pages} {A23} (\bibinfo {year} {2010})}\BibitemShut {NoStop}%
\bibitem [{\citenamefont {Schmit}\ \emph {et~al.}(2010)\citenamefont {Schmit},
  \citenamefont {Dodin},\ and\ \citenamefont {Fisch}}]{Schmit.2010}%
  \BibitemOpen
  \bibfield  {author} {\bibinfo {author} {\bibfnamefont {P.~F.}\ \bibnamefont
  {Schmit}}, \bibinfo {author} {\bibfnamefont {I.~Y.}\ \bibnamefont {Dodin}}, \
  and\ \bibinfo {author} {\bibfnamefont {N.~J.}\ \bibnamefont {Fisch}},\ }\href
  {\doibase 10.1103/PhysRevLett.105.175003} {\bibfield  {journal} {\bibinfo
  {journal} {{P}hys. {R}ev. {L}ett.}\ }\textbf {\bibinfo {volume} {105}},\
  \bibinfo {pages} {175003} (\bibinfo {year} {2010})}\BibitemShut {NoStop}%
\bibitem [{\citenamefont {Grishchuk}(1975)}]{Grishchuk.1975}%
  \BibitemOpen
  \bibfield  {author} {\bibinfo {author} {\bibfnamefont {L.~P.}\ \bibnamefont
  {Grishchuk}},\ }\href {http://jetp.ac.ru/cgi-bin/e/index/r/67/3/p825?a=list}
  {\bibfield  {journal} {\bibinfo  {journal} {{S}ov. {P}hys. {JETP}}\ }\textbf
  {\bibinfo {volume} {67}},\ \bibinfo {pages} {825} (\bibinfo {year}
  {1975})}\BibitemShut {NoStop}%
\bibitem [{\citenamefont {Phan}\ \emph {et~al.}()\citenamefont {Phan},
  \citenamefont {Song},\ and\ \citenamefont {Levitov}}]{Phan2013}%
  \BibitemOpen
  \bibfield  {author} {\bibinfo {author} {\bibfnamefont {T.~V.}\ \bibnamefont
  {Phan}}, \bibinfo {author} {\bibfnamefont {J.~C.~W.}\ \bibnamefont {Song}}, \
  and\ \bibinfo {author} {\bibfnamefont {L.~S.}\ \bibnamefont {Levitov}},\
  }\href {http://arxiv.org/abs/1306.4972} {\enquote {\bibinfo {title}
  {{B}allistic {H}eat {T}ransfer and {E}nergy {W}aves in an {E}lectron
  {S}ystem},}\ }\bibinfo {note} {{a}r{X}iv:1306.4972 (unpublished)}\BibitemShut
  {NoStop}%
\bibitem [{\citenamefont {Briskot}\ \emph {et~al.}(2015)\citenamefont
  {Briskot}, \citenamefont {Sch\"{u}tt}, \citenamefont {Gornyi}, \citenamefont
  {Titov}, \citenamefont {Narozhny},\ and\ \citenamefont
  {Mirlin}}]{Briskot2015}%
  \BibitemOpen
  \bibfield  {author} {\bibinfo {author} {\bibfnamefont {U.}~\bibnamefont
  {Briskot}}, \bibinfo {author} {\bibfnamefont {M.}~\bibnamefont {Sch\"{u}tt}},
  \bibinfo {author} {\bibfnamefont {I.~V.}\ \bibnamefont {Gornyi}}, \bibinfo
  {author} {\bibfnamefont {M.}~\bibnamefont {Titov}}, \bibinfo {author}
  {\bibfnamefont {B.~N.}\ \bibnamefont {Narozhny}}, \ and\ \bibinfo {author}
  {\bibfnamefont {A.~D.}\ \bibnamefont {Mirlin}},\ }\href {\doibase
  10.1103/PhysRevB.92.115426} {\bibfield  {journal} {\bibinfo  {journal}
  {{P}hys. {R}ev. {B}}\ }\textbf {\bibinfo {volume} {92}},\ \bibinfo {pages}
  {115426} (\bibinfo {year} {2015})}\BibitemShut {NoStop}%
\bibitem [{\citenamefont {Ruvalds}(1981)}]{Ruvalds1981}%
  \BibitemOpen
  \bibfield  {author} {\bibinfo {author} {\bibfnamefont {J.}~\bibnamefont
  {Ruvalds}},\ }\href {\doibase 10.1080/00018738100101427} {\bibfield
  {journal} {\bibinfo  {journal} {{A}dv. {P}hys.}\ }\textbf {\bibinfo {volume}
  {30}},\ \bibinfo {pages} {677} (\bibinfo {year} {1981})}\BibitemShut
  {NoStop}%
\bibitem [{\citenamefont {Pinczuk}\ \emph {et~al.}(1981)\citenamefont
  {Pinczuk}, \citenamefont {Shah},\ and\ \citenamefont {Wolff}}]{Pinczuk1981}%
  \BibitemOpen
  \bibfield  {author} {\bibinfo {author} {\bibfnamefont {A.}~\bibnamefont
  {Pinczuk}}, \bibinfo {author} {\bibfnamefont {J.}~\bibnamefont {Shah}}, \
  and\ \bibinfo {author} {\bibfnamefont {P.~A.}\ \bibnamefont {Wolff}},\ }\href
  {\doibase 10.1103/PhysRevLett.47.1487} {\bibfield  {journal} {\bibinfo
  {journal} {{P}hys. {R}ev. {L}ett.}\ }\textbf {\bibinfo {volume} {47}},\
  \bibinfo {pages} {1487} (\bibinfo {year} {1981})}\BibitemShut {NoStop}%
\bibitem [{\citenamefont {Padmanabhan}\ \emph {et~al.}(2014)\citenamefont
  {Padmanabhan}, \citenamefont {Young}, \citenamefont {Henstridge},
  \citenamefont {Bhowmick}, \citenamefont {Bhattacharya},\ and\ \citenamefont
  {Merlin}}]{Padmanabhan2014}%
  \BibitemOpen
  \bibfield  {author} {\bibinfo {author} {\bibfnamefont {P.}~\bibnamefont
  {Padmanabhan}}, \bibinfo {author} {\bibfnamefont {S.~M.}\ \bibnamefont
  {Young}}, \bibinfo {author} {\bibfnamefont {M.}~\bibnamefont {Henstridge}},
  \bibinfo {author} {\bibfnamefont {S.}~\bibnamefont {Bhowmick}}, \bibinfo
  {author} {\bibfnamefont {P.~K.}\ \bibnamefont {Bhattacharya}}, \ and\
  \bibinfo {author} {\bibfnamefont {R.}~\bibnamefont {Merlin}},\ }\href
  {\doibase 10.1103/PhysRevLett.113.027402} {\bibfield  {journal} {\bibinfo
  {journal} {{P}hys. {R}ev. {L}ett.}\ }\textbf {\bibinfo {volume} {113}},\
  \bibinfo {pages} {027402} (\bibinfo {year} {2014})}\BibitemShut {NoStop}%
\bibitem [{\citenamefont {Sunyaev}\ and\ \citenamefont
  {Zeldovich}(1970)}]{Sunyaev1970}%
  \BibitemOpen
  \bibfield  {author} {\bibinfo {author} {\bibfnamefont {R.~A.}\ \bibnamefont
  {Sunyaev}}\ and\ \bibinfo {author} {\bibfnamefont {Y.~B.}\ \bibnamefont
  {Zeldovich}},\ }\href {http://link.springer.com/article/10.1007%2FBF00653471}
  {\bibfield  {journal} {\bibinfo  {journal} {{A}strophys. {S}pace {S}ci.}\
  }\textbf {\bibinfo {volume} {7}},\ \bibinfo {pages} {3} (\bibinfo {year}
  {1970})}\BibitemShut {NoStop}%
\bibitem [{\citenamefont {Stockman}(2004)}]{Stockman2004}%
  \BibitemOpen
  \bibfield  {author} {\bibinfo {author} {\bibfnamefont {M.~I.}\ \bibnamefont
  {Stockman}},\ }\href {\doibase 10.1103/PhysRevLett.93.137404} {\bibfield
  {journal} {\bibinfo  {journal} {{P}hys. {R}ev. {L}ett.}\ }\textbf {\bibinfo
  {volume} {93}},\ \bibinfo {pages} {137404} (\bibinfo {year}
  {2004})}\BibitemShut {NoStop}%
\bibitem [{\citenamefont {Grigorenko}\ \emph {et~al.}(2012)\citenamefont
  {Grigorenko}, \citenamefont {Polini},\ and\ \citenamefont
  {Novoselov}}]{Grigorenko2012}%
  \BibitemOpen
  \bibfield  {author} {\bibinfo {author} {\bibfnamefont {A.~N.}\ \bibnamefont
  {Grigorenko}}, \bibinfo {author} {\bibfnamefont {M.}~\bibnamefont {Polini}},
  \ and\ \bibinfo {author} {\bibfnamefont {K.~S.}\ \bibnamefont {Novoselov}},\
  }\href {\doibase 10.1038/nphoton.2012.262} {\bibfield  {journal} {\bibinfo
  {journal} {{N}at. {P}hotonics}\ }\textbf {\bibinfo {volume} {6}},\ \bibinfo
  {pages} {749} (\bibinfo {year} {2012})}\BibitemShut {NoStop}%
\bibitem [{\citenamefont {Basov}\ \emph {et~al.}(2014)\citenamefont {Basov},
  \citenamefont {Fogler}, \citenamefont {Lanzara}, \citenamefont {Wang},\ and\
  \citenamefont {Zhang}}]{Basov2014}%
  \BibitemOpen
  \bibfield  {author} {\bibinfo {author} {\bibfnamefont {D.}~\bibnamefont
  {Basov}}, \bibinfo {author} {\bibfnamefont {M.}~\bibnamefont {Fogler}},
  \bibinfo {author} {\bibfnamefont {A.}~\bibnamefont {Lanzara}}, \bibinfo
  {author} {\bibfnamefont {F.}~\bibnamefont {Wang}}, \ and\ \bibinfo {author}
  {\bibfnamefont {Y.}~\bibnamefont {Zhang}},\ }\href {\doibase
  10.1103/RevModPhys.86.959} {\bibfield  {journal} {\bibinfo  {journal} {{R}ev.
  {M}od. {P}hys.}\ }\textbf {\bibinfo {volume} {86}},\ \bibinfo {pages} {959}
  (\bibinfo {year} {2014})}\BibitemShut {NoStop}%
\bibitem [{\citenamefont {{Garc\'{\i}a de Abajo}}(2014)}]{GarciadeAbajo2014}%
  \BibitemOpen
  \bibfield  {author} {\bibinfo {author} {\bibfnamefont {F.~J.}\ \bibnamefont
  {{Garc\'{\i}a de Abajo}}},\ }\href {\doibase 10.1021/ph400147y} {\bibfield
  {journal} {\bibinfo  {journal} {{ACS} {P}hotonics}\ }\textbf {\bibinfo
  {volume} {1}},\ \bibinfo {pages} {135} (\bibinfo {year} {2014})}\BibitemShut
  {NoStop}%
\bibitem [{\citenamefont {Vafek}(2006)}]{Vafek2006}%
  \BibitemOpen
  \bibfield  {author} {\bibinfo {author} {\bibfnamefont {O.}~\bibnamefont
  {Vafek}},\ }\href {\doibase 10.1103/PhysRevLett.97.266406} {\bibfield
  {journal} {\bibinfo  {journal} {{P}hys. {R}ev. {L}ett.}\ }\textbf {\bibinfo
  {volume} {97}},\ \bibinfo {pages} {266406} (\bibinfo {year}
  {2006})}\BibitemShut {NoStop}%
\bibitem [{\citenamefont {Svintsov}\ \emph {et~al.}(2012)\citenamefont
  {Svintsov}, \citenamefont {Vyurkov}, \citenamefont {Yurchenko}, \citenamefont
  {Otsuji},\ and\ \citenamefont {Ryzhii}}]{Svintsov2012}%
  \BibitemOpen
  \bibfield  {author} {\bibinfo {author} {\bibfnamefont {D.}~\bibnamefont
  {Svintsov}}, \bibinfo {author} {\bibfnamefont {V.}~\bibnamefont {Vyurkov}},
  \bibinfo {author} {\bibfnamefont {S.}~\bibnamefont {Yurchenko}}, \bibinfo
  {author} {\bibfnamefont {T.}~\bibnamefont {Otsuji}}, \ and\ \bibinfo {author}
  {\bibfnamefont {V.}~\bibnamefont {Ryzhii}},\ }\href {\doibase
  10.1063/1.4705382} {\bibfield  {journal} {\bibinfo  {journal} {{J}. {A}ppl.
  {P}hys.}\ }\textbf {\bibinfo {volume} {111}},\ \bibinfo {pages} {083715}
  (\bibinfo {year} {2012})}\BibitemShut {NoStop}%
\bibitem [{\citenamefont {Kashuba}(2008)}]{Kashuba2008}%
  \BibitemOpen
  \bibfield  {author} {\bibinfo {author} {\bibfnamefont {A.~B.}\ \bibnamefont
  {Kashuba}},\ }\href {\doibase 10.1103/PhysRevB.78.085415} {\bibfield
  {journal} {\bibinfo  {journal} {{P}hys. {R}ev. {B}}\ }\textbf {\bibinfo
  {volume} {78}},\ \bibinfo {pages} {085415} (\bibinfo {year}
  {2008})}\BibitemShut {NoStop}%
\bibitem [{\citenamefont {M\"{u}ller}\ \emph {et~al.}(2008)\citenamefont
  {M\"{u}ller}, \citenamefont {Fritz},\ and\ \citenamefont
  {Sachdev}}]{Muller2008a}%
  \BibitemOpen
  \bibfield  {author} {\bibinfo {author} {\bibfnamefont {M.}~\bibnamefont
  {M\"{u}ller}}, \bibinfo {author} {\bibfnamefont {L.}~\bibnamefont {Fritz}}, \
  and\ \bibinfo {author} {\bibfnamefont {S.}~\bibnamefont {Sachdev}},\ }\href
  {\doibase 10.1103/PhysRevB.78.115406} {\bibfield  {journal} {\bibinfo
  {journal} {{P}hys. {R}ev. {B}}\ }\textbf {\bibinfo {volume} {78}},\ \bibinfo
  {pages} {115406} (\bibinfo {year} {2008})}\BibitemShut {NoStop}%
\bibitem [{\citenamefont {Sch\"{u}tt}\ \emph {et~al.}(2011)\citenamefont
  {Sch\"{u}tt}, \citenamefont {Ostrovsky}, \citenamefont {Gornyi},\ and\
  \citenamefont {Mirlin}}]{Schuett2011}%
  \BibitemOpen
  \bibfield  {author} {\bibinfo {author} {\bibfnamefont {M.}~\bibnamefont
  {Sch\"{u}tt}}, \bibinfo {author} {\bibfnamefont {P.~M.}\ \bibnamefont
  {Ostrovsky}}, \bibinfo {author} {\bibfnamefont {I.~V.}\ \bibnamefont
  {Gornyi}}, \ and\ \bibinfo {author} {\bibfnamefont {A.~D.}\ \bibnamefont
  {Mirlin}},\ }\href {\doibase 10.1103/PhysRevB.83.155441} {\bibfield
  {journal} {\bibinfo  {journal} {{P}hys. {R}ev. {B}}\ }\textbf {\bibinfo
  {volume} {83}},\ \bibinfo {pages} {155441} (\bibinfo {year}
  {2011})}\BibitemShut {NoStop}%
\bibitem [{\citenamefont {Orenstein}\ and\ \citenamefont
  {Dodge}(2015)}]{Orenstein2015}%
  \BibitemOpen
  \bibfield  {author} {\bibinfo {author} {\bibfnamefont {J.}~\bibnamefont
  {Orenstein}}\ and\ \bibinfo {author} {\bibfnamefont {J.~S.}\ \bibnamefont
  {Dodge}},\ }\href {\doibase 10.1103/PhysRevB.92.134507} {\bibfield  {journal}
  {\bibinfo  {journal} {{P}hys. {R}ev. {B}}\ }\textbf {\bibinfo {volume}
  {92}},\ \bibinfo {pages} {134507} (\bibinfo {year} {2015})}\BibitemShut
  {NoStop}%
\bibitem [{SM()}]{SM}%
  \BibitemOpen
  \href@noop {} {}\bibinfo {note} {See {S}upplemental {M}aterial at [{URL} to
  be inserted by publisher] for technical details.}\BibitemShut {Stop}%
\bibitem [{\citenamefont {Massicotte}\ \emph {et~al.}(2016)\citenamefont
  {Massicotte}, \citenamefont {Schmidt}, \citenamefont {Vialla}, \citenamefont
  {Watanabe}, \citenamefont {Taniguchi}, \citenamefont {Tielrooij},\ and\
  \citenamefont {Koppens}}]{Massicotte2016}%
  \BibitemOpen
  \bibfield  {author} {\bibinfo {author} {\bibfnamefont {M.}~\bibnamefont
  {Massicotte}}, \bibinfo {author} {\bibfnamefont {P.}~\bibnamefont {Schmidt}},
  \bibinfo {author} {\bibfnamefont {F.}~\bibnamefont {Vialla}}, \bibinfo
  {author} {\bibfnamefont {K.}~\bibnamefont {Watanabe}}, \bibinfo {author}
  {\bibfnamefont {T.}~\bibnamefont {Taniguchi}}, \bibinfo {author}
  {\bibfnamefont {K.~J.}\ \bibnamefont {Tielrooij}}, \ and\ \bibinfo {author}
  {\bibfnamefont {F.~H.~L.}\ \bibnamefont {Koppens}},\ }\href
  {http://dx.doi.org/10.1038/ncomms12174} {\bibfield  {journal} {\bibinfo
  {journal} {Nat Commun}\ }\textbf {\bibinfo {volume} {7}},\  (\bibinfo {year}
  {2016})}\BibitemShut {NoStop}%
\bibitem [{\citenamefont {Fang}\ \emph {et~al.}(2012)\citenamefont {Fang},
  \citenamefont {Wang}, \citenamefont {Liu}, \citenamefont {Schlather},
  \citenamefont {Ajayan}, \citenamefont {Koppens}, \citenamefont {Nordlander},\
  and\ \citenamefont {Halas}}]{Fang2012pid}%
  \BibitemOpen
  \bibfield  {author} {\bibinfo {author} {\bibfnamefont {Z.}~\bibnamefont
  {Fang}}, \bibinfo {author} {\bibfnamefont {Y.}~\bibnamefont {Wang}}, \bibinfo
  {author} {\bibfnamefont {Z.}~\bibnamefont {Liu}}, \bibinfo {author}
  {\bibfnamefont {A.}~\bibnamefont {Schlather}}, \bibinfo {author}
  {\bibfnamefont {P.~M.}\ \bibnamefont {Ajayan}}, \bibinfo {author}
  {\bibfnamefont {F.~H.~L.}\ \bibnamefont {Koppens}}, \bibinfo {author}
  {\bibfnamefont {P.}~\bibnamefont {Nordlander}}, \ and\ \bibinfo {author}
  {\bibfnamefont {N.~J.}\ \bibnamefont {Halas}},\ }\href {\doibase
  10.1021/nn304028b} {\bibfield  {journal} {\bibinfo  {journal} {ACS Nano}\
  }\textbf {\bibinfo {volume} {6}},\ \bibinfo {pages} {10222} (\bibinfo {year}
  {2012})}\BibitemShut {NoStop}%
\bibitem [{\citenamefont {Huber}\ \emph {et~al.}(2001)\citenamefont {Huber},
  \citenamefont {Tauser}, \citenamefont {Brodschelm}, \citenamefont {Bichler},
  \citenamefont {Abstreiter},\ and\ \citenamefont {Leitenstorfer}}]{Huber2001}%
  \BibitemOpen
  \bibfield  {author} {\bibinfo {author} {\bibfnamefont {R.}~\bibnamefont
  {Huber}}, \bibinfo {author} {\bibfnamefont {F.}~\bibnamefont {Tauser}},
  \bibinfo {author} {\bibfnamefont {A.}~\bibnamefont {Brodschelm}}, \bibinfo
  {author} {\bibfnamefont {M.}~\bibnamefont {Bichler}}, \bibinfo {author}
  {\bibfnamefont {G.}~\bibnamefont {Abstreiter}}, \ and\ \bibinfo {author}
  {\bibfnamefont {A.}~\bibnamefont {Leitenstorfer}},\ }\href {\doibase
  10.1038/35104522} {\bibfield  {journal} {\bibinfo  {journal} {{N}ature}\
  }\textbf {\bibinfo {volume} {414}},\ \bibinfo {pages} {286} (\bibinfo {year}
  {2001})}\BibitemShut {NoStop}%
\bibitem [{\citenamefont {Ma}\ \emph {et~al.}(2016)\citenamefont {Ma},
  \citenamefont {Andersen}, \citenamefont {Nair}, \citenamefont {Gabor},
  \citenamefont {Massicotte}, \citenamefont {Lui}, \citenamefont {Young},
  \citenamefont {Fang}, \citenamefont {Watanabe}, \citenamefont {Taniguchi},
  \citenamefont {Kong}, \citenamefont {Gedik}, \citenamefont {Koppens},\ and\
  \citenamefont {Jarillo-Herrero}}]{Ma2016}%
  \BibitemOpen
  \bibfield  {author} {\bibinfo {author} {\bibfnamefont {Q.}~\bibnamefont
  {Ma}}, \bibinfo {author} {\bibfnamefont {T.~I.}\ \bibnamefont {Andersen}},
  \bibinfo {author} {\bibfnamefont {N.~L.}\ \bibnamefont {Nair}}, \bibinfo
  {author} {\bibfnamefont {N.~M.}\ \bibnamefont {Gabor}}, \bibinfo {author}
  {\bibfnamefont {M.}~\bibnamefont {Massicotte}}, \bibinfo {author}
  {\bibfnamefont {C.~H.}\ \bibnamefont {Lui}}, \bibinfo {author} {\bibfnamefont
  {A.~F.}\ \bibnamefont {Young}}, \bibinfo {author} {\bibfnamefont
  {W.}~\bibnamefont {Fang}}, \bibinfo {author} {\bibfnamefont {K.}~\bibnamefont
  {Watanabe}}, \bibinfo {author} {\bibfnamefont {T.}~\bibnamefont {Taniguchi}},
  \bibinfo {author} {\bibfnamefont {J.}~\bibnamefont {Kong}}, \bibinfo {author}
  {\bibfnamefont {N.}~\bibnamefont {Gedik}}, \bibinfo {author} {\bibfnamefont
  {F.~H.~L.}\ \bibnamefont {Koppens}}, \ and\ \bibinfo {author} {\bibfnamefont
  {P.}~\bibnamefont {Jarillo-Herrero}},\ }\href {\doibase 10.1038/nphys3620}
  {\bibfield  {journal} {\bibinfo  {journal} {{N}ature {P}hys.}\ }\textbf
  {\bibinfo {volume} {12}},\ \bibinfo {pages} {455} (\bibinfo {year}
  {2016})}\BibitemShut {NoStop}%
\bibitem [{\citenamefont {Alonso-Gonz\'{a}lez}\ \emph {et~al.}()\citenamefont
  {Alonso-Gonz\'{a}lez}, \citenamefont {Nikitin}, \citenamefont {Gao},
  \citenamefont {Woessner}, \citenamefont {Lundeberg}, \citenamefont
  {Principi}, \citenamefont {Forcellini}, \citenamefont {Yan}, \citenamefont
  {Velez}, \citenamefont {Huber}, \citenamefont {Watanabe}, \citenamefont
  {Taniguchi}, \citenamefont {Hueso}, \citenamefont {Polini}, \citenamefont
  {Hone}, \citenamefont {Koppens},\ and\ \citenamefont
  {Hillenbrand}}]{Alonso-Gonzalez2016}%
  \BibitemOpen
  \bibfield  {author} {\bibinfo {author} {\bibfnamefont {P.}~\bibnamefont
  {Alonso-Gonz\'{a}lez}}, \bibinfo {author} {\bibfnamefont {A.~Y.}\
  \bibnamefont {Nikitin}}, \bibinfo {author} {\bibfnamefont {Y.}~\bibnamefont
  {Gao}}, \bibinfo {author} {\bibfnamefont {A.}~\bibnamefont {Woessner}},
  \bibinfo {author} {\bibfnamefont {M.~B.}\ \bibnamefont {Lundeberg}}, \bibinfo
  {author} {\bibfnamefont {A.}~\bibnamefont {Principi}}, \bibinfo {author}
  {\bibfnamefont {N.}~\bibnamefont {Forcellini}}, \bibinfo {author}
  {\bibfnamefont {W.}~\bibnamefont {Yan}}, \bibinfo {author} {\bibfnamefont
  {S.}~\bibnamefont {Velez}}, \bibinfo {author} {\bibfnamefont {A.~J.}\
  \bibnamefont {Huber}}, \bibinfo {author} {\bibfnamefont {K.}~\bibnamefont
  {Watanabe}}, \bibinfo {author} {\bibfnamefont {T.}~\bibnamefont {Taniguchi}},
  \bibinfo {author} {\bibfnamefont {L.~E.}\ \bibnamefont {Hueso}}, \bibinfo
  {author} {\bibfnamefont {M.}~\bibnamefont {Polini}}, \bibinfo {author}
  {\bibfnamefont {J.}~\bibnamefont {Hone}}, \bibinfo {author} {\bibfnamefont
  {F.~H.~L.}\ \bibnamefont {Koppens}}, \ and\ \bibinfo {author} {\bibfnamefont
  {R.}~\bibnamefont {Hillenbrand}},\ }\href {http://arxiv.org/abs/1601.05753}
  {\enquote {\bibinfo {title} {{U}ltra-confined acoustic {TH}z graphene
  plasmons revealed by photocurrent nanoscopy},}\ }\bibinfo {note}
  {{a}r{X}iv:1601.05753 (unpublished)}\BibitemShut {NoStop}%
\bibitem [{\citenamefont {Buisson}\ \emph {et~al.}(1994)\citenamefont
  {Buisson}, \citenamefont {Xavier},\ and\ \citenamefont
  {Richard}}]{Buisson1994}%
  \BibitemOpen
  \bibfield  {author} {\bibinfo {author} {\bibfnamefont {O.}~\bibnamefont
  {Buisson}}, \bibinfo {author} {\bibfnamefont {P.}~\bibnamefont {Xavier}}, \
  and\ \bibinfo {author} {\bibfnamefont {J.}~\bibnamefont {Richard}},\ }\href
  {\doibase 10.1103/PhysRevLett.73.3153} {\bibfield  {journal} {\bibinfo
  {journal} {{P}hys. {R}ev. {L}ett.}\ }\textbf {\bibinfo {volume} {73}},\
  \bibinfo {pages} {3153} (\bibinfo {year} {1994})}\BibitemShut {NoStop}%
\bibitem [{\citenamefont {Stinson}\ \emph {et~al.}(2014)\citenamefont
  {Stinson}, \citenamefont {Wu}, \citenamefont {Jiang}, \citenamefont {Fei},
  \citenamefont {Rodin}, \citenamefont {Chapler}, \citenamefont {McLeod},
  \citenamefont {Castro~Neto}, \citenamefont {Lee}, \citenamefont {Fogler},\
  and\ \citenamefont {Basov}}]{Stinson2014}%
  \BibitemOpen
  \bibfield  {author} {\bibinfo {author} {\bibfnamefont {H.~T.}\ \bibnamefont
  {Stinson}}, \bibinfo {author} {\bibfnamefont {J.~S.}\ \bibnamefont {Wu}},
  \bibinfo {author} {\bibfnamefont {B.~Y.}\ \bibnamefont {Jiang}}, \bibinfo
  {author} {\bibfnamefont {Z.}~\bibnamefont {Fei}}, \bibinfo {author}
  {\bibfnamefont {A.~S.}\ \bibnamefont {Rodin}}, \bibinfo {author}
  {\bibfnamefont {B.~C.}\ \bibnamefont {Chapler}}, \bibinfo {author}
  {\bibfnamefont {A.~S.}\ \bibnamefont {McLeod}}, \bibinfo {author}
  {\bibfnamefont {A.}~\bibnamefont {Castro~Neto}}, \bibinfo {author}
  {\bibfnamefont {Y.~S.}\ \bibnamefont {Lee}}, \bibinfo {author} {\bibfnamefont
  {M.~M.}\ \bibnamefont {Fogler}}, \ and\ \bibinfo {author} {\bibfnamefont
  {D.~N.}\ \bibnamefont {Basov}},\ }\href {\doibase 10.1103/PhysRevB.90.014502}
  {\bibfield  {journal} {\bibinfo  {journal} {{P}hys. {R}ev. {B}}\ }\textbf
  {\bibinfo {volume} {90}},\ \bibinfo {pages} {014502} (\bibinfo {year}
  {2014})}\BibitemShut {NoStop}%
\bibitem [{Sup()}]{Supplementary}%
  \BibitemOpen
  \href@noop {} {}\bibinfo {note} {See Supplemental Material [{URL} to be
  inserted by publisher] , which includes Refs. [45-47].}\BibitemShut {Stop}%
\bibitem [{\citenamefont {M\"{u}ller}\ and\ \citenamefont
  {Sachdev}(2008)}]{Muller2008}%
  \BibitemOpen
  \bibfield  {author} {\bibinfo {author} {\bibfnamefont {M.}~\bibnamefont
  {M\"{u}ller}}\ and\ \bibinfo {author} {\bibfnamefont {S.}~\bibnamefont
  {Sachdev}},\ }\href {\doibase 10.1103/PhysRevB.78.115419} {\bibfield
  {journal} {\bibinfo  {journal} {{P}hys. {R}ev. {B}}\ }\textbf {\bibinfo
  {volume} {78}},\ \bibinfo {pages} {115419} (\bibinfo {year}
  {2008})}\BibitemShut {NoStop}%
\bibitem [{\citenamefont {M\"{u}ller}\ \emph {et~al.}(2009)\citenamefont
  {M\"{u}ller}, \citenamefont {Schmalian},\ and\ \citenamefont
  {Fritz}}]{Muller2009}%
  \BibitemOpen
  \bibfield  {author} {\bibinfo {author} {\bibfnamefont {M.}~\bibnamefont
  {M\"{u}ller}}, \bibinfo {author} {\bibfnamefont {J.}~\bibnamefont
  {Schmalian}}, \ and\ \bibinfo {author} {\bibfnamefont {L.}~\bibnamefont
  {Fritz}},\ }\href {\doibase 10.1103/PhysRevLett.103.025301} {\bibfield
  {journal} {\bibinfo  {journal} {{P}hys. {R}ev. {L}ett.}\ }\textbf {\bibinfo
  {volume} {103}},\ \bibinfo {pages} {025301} (\bibinfo {year}
  {2009})}\BibitemShut {NoStop}%
\bibitem [{\citenamefont {Narozhny}\ \emph {et~al.}(2015)\citenamefont
  {Narozhny}, \citenamefont {Gornyi}, \citenamefont {Titov}, \citenamefont
  {Sch\"utt},\ and\ \citenamefont {Mirlin}}]{Narozhny.2015}%
  \BibitemOpen
  \bibfield  {author} {\bibinfo {author} {\bibfnamefont {B.~N.}\ \bibnamefont
  {Narozhny}}, \bibinfo {author} {\bibfnamefont {I.~V.}\ \bibnamefont
  {Gornyi}}, \bibinfo {author} {\bibfnamefont {M.}~\bibnamefont {Titov}},
  \bibinfo {author} {\bibfnamefont {M.}~\bibnamefont {Sch\"utt}}, \ and\
  \bibinfo {author} {\bibfnamefont {A.~D.}\ \bibnamefont {Mirlin}},\ }\href
  {\doibase 10.1103/PhysRevB.91.035414} {\bibfield  {journal} {\bibinfo
  {journal} {{P}hys. {R}ev. {B}}\ }\textbf {\bibinfo {volume} {91}},\ \bibinfo
  {pages} {035414} (\bibinfo {year} {2015})}\BibitemShut {NoStop}%
\end{thebibliography}%

\clearpage

\begin{center}
	\textbf{Supplementary material for ``Adiabatic amplification of plasmons and demons in 2D systems''}
\end{center}
\setcounter{equation}{0}
\setcounter{figure}{0}
\setcounter{table}{0}
\setcounter{page}{1}
	
	\section{Conductivity kernel}
	\label{sec:Drude}
	
	The electric current induced in a system by a weak external field can be derived from the Boltzmann transport equation.
	Consider a general 2D system consisting of $N$ layers.
	Let $f^{(i)}(\mathbf{k}, t)$ be the electron distribution function of layer $i$.
	The total
	distribution function $f(\mathbf{k})$ is the sum
	$f(\mathbf{k}, t) = \sum_{i = 1}^N f^{(i)}(\mathbf{k}, t)$.
	Function $f(\mathbf{k})$ can be decomposed into partial waves of different angular momenta.
	When the external electric field is uniform and the unperturbed Hamiltonian
	is isotropic, only the $s$- and $p$-waves contribute:
	$f(\mathbf{k}) =  f_s(\mathbf{k}) + f_p(\mathbf{k})$.
	The angle-independent part $f_s$ contains the information about the nonequlibrum process, e.g., the change of electron temperature $T(t)$ and chemical potential $\mu(t)$.
	The $p$-wave part $f_p$ determines the total electric current of the system. In the linear-response regime
	$f_p$ is a small correction to $f_s$.
	The linearized Boltzmann equation has the form
	\begin{align}
		\partial_t f - e \mathbf{E}\, \partial_\mathbf{k} f 
		= -\gamma_0 [f_s - f_0(t)] - \gamma(t) f_p  \,,
		\label{eqn:boltzmann}
	\end{align}
	where $\gamma_0$ is the relaxation rate of the $s$-wave part $f_s$,
	function $f_0(t)$ is the Fermi-Dirac distribution defined by $T(t)$, $\mu(t)$, and $\gamma(t)$ is the relaxation rate of the $p$-wave part due to disorder, electron-phonon, and  electron-electron scattering.
	Keeping the leading-order terms in the external field,
	we obtain separate equations for the two partial waves:
	\begin{align}
		\partial_t f_s &= -\gamma_0 [f_s - f_0(t)]\,,  
		\label{eqn:channelss}\\
		\partial_t f_p  &=  e \mathbf{E}\, \partial_\mathbf{k} f_s  - \gamma(t) f_p\,. 
		\label{eqn:channelsp}
	\end{align}
	If $\gamma_0$ is the fastest rate in the problem,
	the approximate solutions of these equations are
	$f_s = f_0(t)$ and
	\begin{align}
		f_p (t) = \int\limits_{-\infty}^{t}
		e \mathbf{E}(t_0) \partial_\mathbf{k} f_0(t_0)
		e^{-s(t,t_0)} d t_0 \,,
	\end{align}
	where $s(t, t_0) = \int_{t_0}^{t} \gamma(t^\prime) d t^{\prime}$
	is the accumulated damping exponent.
	The current is given by
	\[
	\textbf{j}(t) 
	= -e \sum_{\mathbf{k}} \mathbf{v}(\mathbf{k}, t) f_p(\mathbf{k},t)
	\equiv
	\int\limits_{-\infty}^{\infty} \sigma(t,t_0) \mathbf{E}(t_0) d t_0 \,,
	\label{eqn:current_definition}
	\]
	where
	\begin{align}
		\sigma(t,t_0) &= \frac{1}{\pi}\, \mathcal{D}(t,t_0) \theta(t - t_0)\,,
		\\
		\mathcal{D}(t,t_0) &= -\frac{\pi}{2}\,  \sum_{\mathbf{k}}
		e^2\mathbf{v}(\mathbf{k},t) \partial_{\mathbf{k}} f_0(t_0)
		e^{-s(t,t_0)},
		\label{eqn:D_t_t_0}\\
		\mathbf{v}(\mathbf{k}, t) &=  [\partial_{\mathbf{k}} f_0(\mathbf{k},t)]^{-1}
		\sum_{i = 1}^N \mathbf{v}^{(i)}(\mathbf{k})
		\partial_{\mathbf{k}} f^{(i)}_0(\mathbf{k},t)\,.
		\label{eqn:v}
	\end{align}
	We define the instantaneous Drude weight $D(t_0)$ from
	the condition that the current generated at time $t_0 + 0$ by the electric field $\mathbf{E}(t_0)$ equals $\frac{1}{\pi} D(t_0) \mathbf{E}(t_0)$.
	Combined with the definition (Eq.~(7) of the main text) of $\kappa(t)$, this implies
	\begin{equation}
		\mathcal{D}(t_0, t_0) = D(t_0) = 
		D(0) e^{-2 \int_{0}^{t_0} \kappa(t') d t'}\,.
		\label{eqn:D}
	\end{equation}

	Let us first examine a single-layer system, $N = 1$, where the average quasiparticle velocity $\mathbf{v}(\mathbf{k}, t)$ is simply
	$\mathbf{v}^{(1)}(\mathbf{k})$ and
	does not change with $t$.
	Suppose this system is cooling after photoexcitation by emitting acoustic phonons.
	A common wisdom is that the phonon emission is much more effective in relaxing the momentum distribution of the electrons than in cooling them.
	Indeed, at $T \gg \mu$, Pauli blocking of the final electron states is unimportant and the typical momentum of an emitted phonon is of the order of the electron momentum $k \sim T / \hbar v_F$.
	On the other hand, the energy of such a phonon $T v_{ph} / v_F$ is much smaller than the typical electron energy $T$ because the sound velocity $v_{ph}$ is much smaller than the Fermi velocity $v_F$.
	Accordingly, $\gamma$ should be significantly larger than the cooling rate $2\kappa$.
	It is then sensible to split $\gamma$ as follows:
	\begin{equation}
		\gamma = \Gamma + 2\kappa\,.
		\label{eqn:gamma_split}
	\end{equation}
	We can transform the two-time Drude weight to
	\begin{align}
		\begin{split}
			\mathcal{D}(t, t_0) &= D(t_0) e^{-s(t, t_0)}\\
			&= D(t_0)
			e^{-2 \int_{t_0}^{t} \kappa(t') d t' - \int_{t_0}^{t} \Gamma(t') d t'} .
		\end{split}\\
		\mathcal{D}(t, t_0) &= D(t) e^{-\int_{t_0}^{t} \Gamma(t') d t'},
		\label{eqn:D_t}
	\end{align}
	which yields Eq.~(6) of the main text.
	
	Consider next a multilayer system where electrons can tunnel between adjacent layers.
	If the tunneling is the major mechanism affecting the electron distribution, and electron momentum $\mathbf{k}$ is conserved in tunneling, then
	it is more natural to
	set $\gamma = \Gamma$ instead of Eq.~\eqref{eqn:gamma_split}.
	The total distribution function
	$f_0 = \sum_{i} f^{(i)}_0$ does not depend on $t_0$.
	However,
	if the effective carrier mass of the layers is different,
	the layer-averaged quasiparticle velocity $\mathbf{v}(\mathbf{k}, t)$ [Eq.~\eqref{eqn:v}] changes with $t$.
	The result for $\mathcal{D}(t, t_0)$ can be written in the form of Eq.~\eqref{eqn:D_t}.
	
	In theory, there is still another case.
	When the electron system is heated due to absorption of photons,
	its energy (or temperature) increases while its momentum remains unchanged.
	If the situation is possible where this photoexcitation does not activate other degrees of freedom (for example, phonons) that cause rapid momentum relaxation,
	then we may set $\gamma = \Gamma$,
	as in the case of tunneling.
	However, the two-time Drude weight $\mathcal{D}(t, t_0)$ increases with time due to $f_0(t_0)$,
	as in the very first case considered.
	As a result, we get
	\begin{align}
		\mathcal{D}(t, t_0) = D(t_0) e^{-\int_{t_0}^{t} \Gamma(t') d t'},
	\end{align}
	in contrast to Eq.~\eqref{eqn:D_t}.
	
	\section{Equation of motion for plasmons}
	If the system has the spatial translational symmetry, different momenta are decoupled.
	For a given $\mathbf{q}$,
	in the linear-response regime, the plasmon dynamics is determined by the electrical conductivity operator $\hat{\sigma}_q$ with the kernel $\sigma_q(t, t_0)$,
	which relates the Fourier components of the current $\mathbf{j}_q$ and the electric field $\mathbf{E}_q$:
	\begin{equation}
		\mathbf{j}_q = \hat{\sigma}_q \mathbf{E}_q
		\equiv \int\limits_{-\infty}^{\infty} \sigma_q(t, t_0) \mathbf{E}_q(t_0) d t_0\,.
		\label{eqn:j_E}
	\end{equation}
	In turn, the charge density $\rho_q$, the electric potential $\Phi_q$, and the current $\mathbf{j}_q$ are related by the Coulomb law and the continuity equation:
	\begin{align}
		\Phi_q = v_q \rho_q\,,\:
		\mathbf{E}_q = -i \mathbf{q} \Phi_q\,,\:
		\partial_{t}\rho = -i \mathbf{q}\, \mathbf{j}_q\,,
		\label{eqn:rho_E_j}
	\end{align}
	where all the quantities are functions of time and
	\begin{equation}
		v_q = 2\pi / \epsilon q
		\label{eqn:v_q}
	\end{equation}
	is the Fourier transform of the bare Coulomb potential screened by the dielectric environment.
	(Note that screening by the electrons themselves should not be included in $\epsilon$ because $\mathbf{E}_{\mathbf{q}}$ is the total electric field.)
	Equations~\eqref{eqn:j_E} and \eqref{eqn:rho_E_j} entail
	\begin{align}
		\partial_t \rho_q + q^2  v_q \hat{\sigma}_q \rho_q =0 \,,
		\label{eqn:density_equation}
	\end{align}
	which is the same as
	\begin{align}
		\partial_t \rho_q
		+ q^2  v_q \int\limits_{-\infty}^{\infty} \sigma_q(t,t_0)
		\rho_q(t_0) dt_0 =0 \,.
		\label{eqn:density_equation_II}
	\end{align}
	With the following model for the conductivity kernel:
	\begin{align}
		\sigma_q(t, t_0) &= \frac{1}{\pi}\, D(t)
		e^{-\Gamma (t - t_0)} \theta(t - t_0) \,,
		\label{eqn:sigma_q}\\
		D(t) &= D(0) e^{-2 \int_{0}^{t} \kappa(t') d t'} \,,
		\label{eqn:2kappa}
	\end{align}
	and taking the time derivative of Eq.~\eqref{eqn:density_equation_II},
	and combining it with Eq.~\eqref{eqn:sigma_q}, we get
	\begin{align}
		\left[ \partial^2_t + \gamma(t) \partial_t + \omega_q^2(t) \right] \rho_q =0 \,,
		\label{eqn:final_density_equation}
	\end{align}
	with
	\begin{gather}
		\gamma = 2\kappa + \Gamma\,,
		\label{eqn:gamma}
	\end{gather}
	which is the result announced in the main text.
	
	\section{Equation of motion for energy waves (demons)}
	\label{sec:hydrodynamic}
	
	\begin{figure}[b]
		\includegraphics[width=2.6in]{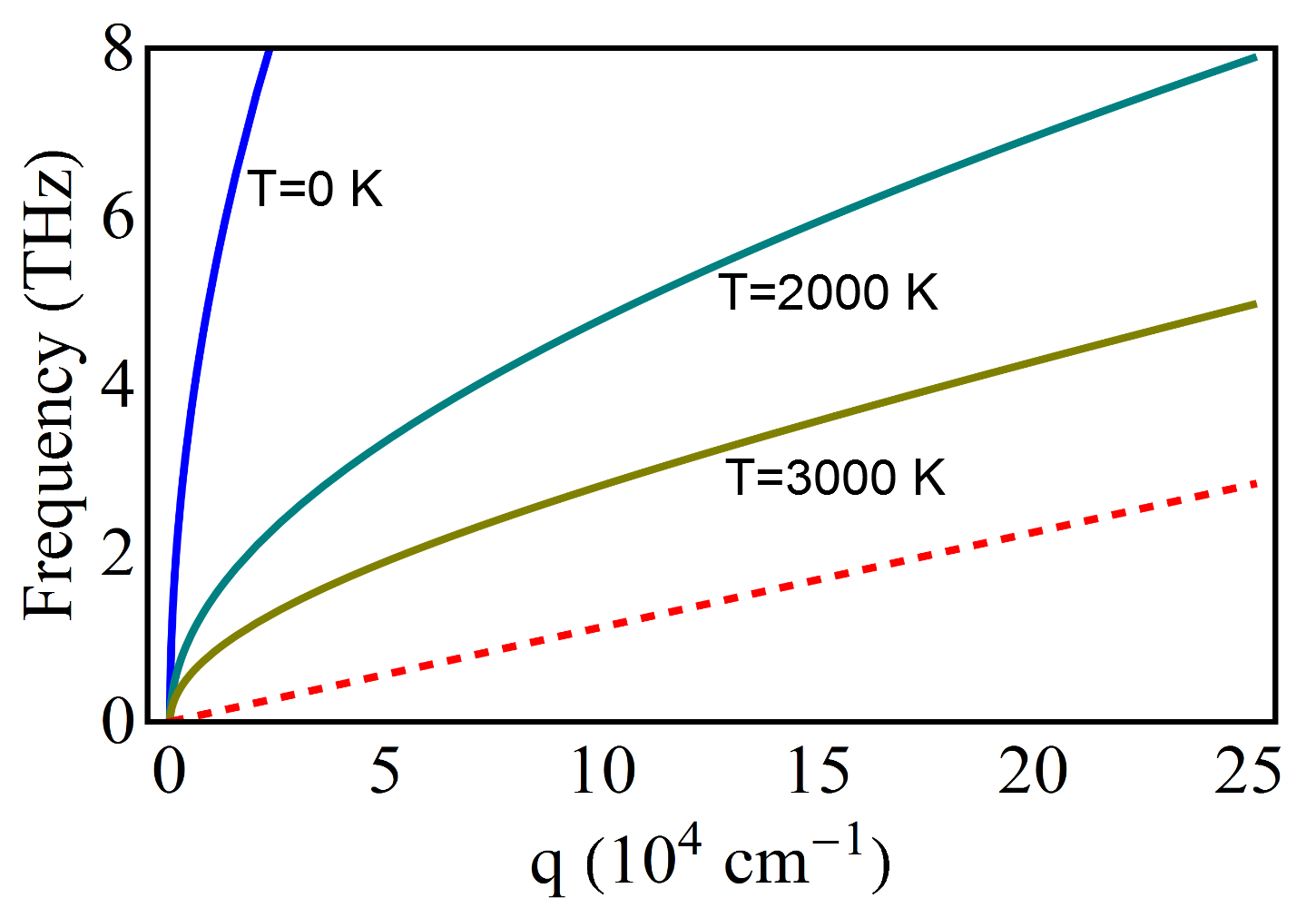} 
		\caption{Dispersion of the energy wave (demon) in graphene at different $T$ calculated from Eqs.~\eqref{eqn:n}, \eqref{eqn:n_E}, and \eqref{eqn:energy_wave_disperison}.
			The red dashed line is the infinite temperature limit, $\omega = \frac{1}{\sqrt{2}}\, v_F q$. The electron concentration is $n = 2.0 \times 10^{12} \unit{cm}^{-2}$ and the dielectric constant is $\epsilon = 1$. }
		\label{fig:energy_wave_dispersion}
	\end{figure}
	
	The linearized hydrodynamic equations~\cite{Muller2008,Muller2008a, Briskot2015,Phan2013,Muller2009,Narozhny.2015} (in the notations of Ref.~\onlinecite{Briskot2015}) are
	\begin{align}
		& \partial_t n + {\nabla} (n \mathbf{u})= -\frac{1}{e}\, \sigma_Q {\nabla}\left(\mathbf{E} + \frac{1}{e} T\, {\nabla} \frac{\mu}{T}\right),
		\label{eqn:hydro1}\\
		& \partial_t n_E + {\nabla} \mathbf{j}_E= 0 \,,
		\label{eqn:hydro2}\\
		& \partial_t \mathbf{j}_E + v_F^2 {\nabla} P = -e n v^2_F \mathbf{E}
		+ \eta {\nabla}^2 \mathbf{u} + \zeta {\nabla} ({\nabla} \mathbf{u}), 
		\label{eqn:hydro3}
	\end{align}
	where $\mathbf{j}_E \equiv \langle\varepsilon(\mathbf{k}) \mathbf{v}(\mathbf{k})\rangle = (n_E + P) \mathbf{u}$ is the energy current,
	$P = \frac12 n_E$ is the pressure,
	$\eta$ and $\zeta$ are the shear and bulk viscosities,
	$\sigma_Q$ is the conductivity,
	and the angular brackets mean the integral of a quantity over electron momenta $\mathbf{k}$ with the weight equal to the shifted Fermi distribution function
	$f = f_0(\mu, T, \epsilon - \mathbf{k}\mathbf{u})$. 
	In equilibrium [$\mathbf{u} \equiv 0$; $\mu(\mathbf{r}, t), T(\mathbf{r}, t) = \mathrm{const}$] the electron concentration $n$ and energy density $n_E$
	have the following analytical form
	\begin{align}
		n &= \int\limits_{-\infty}^{\infty} \left[ f_0(\mu,T,\epsilon) - f_0(0, 0,\epsilon) \right] g(\epsilon) d\epsilon
		\notag \\
		&= \frac{2}{\pi} \frac{T^2}{\hbar^2 v_F^2} \Bigl[\frac{\pi^2}{6}
		+ \frac{1}{2}
		\frac{\mu^2}{T^2}
		+ 2\, \mathrm{Li}_2\bigl(-e^{-\mu/ T}\bigl) \Bigr] ,
		\label{eqn:n}\\
		n_E &= \int\limits_{-\infty}^{\infty} \left[ f_0(\mu,T,\epsilon) - f_0(0, 0,\epsilon)\right] \epsilon g(\epsilon) d\epsilon
		\notag \\
		&= \frac{2}{\pi} \frac{T^3}{\hbar^2 v_F^2} \Bigl[
		\frac{\pi^2}{3}\, \frac{\mu}{T} + \frac{1}{3}\, \frac{\mu^3}{T^3} 
		-4\, \mathrm{Li}_3 \bigl(-e^{-\mu/T}\bigr) \Bigr] .
		\label{eqn:n_E}
	\end{align}
	In these expression, $g(\epsilon) = ({2} / {\pi}) (\,|\epsilon| / \hbar^2 v_F^2)$ is the electron density of states of graphene,
	$\mathrm{Li}_z(x)$ is the polylogarithm function,
	and $f_0(0, 0, \epsilon) = \Theta(-\epsilon)$,
	where
	$\Theta(x)$ is the unit step function.
	For weakly doped graphene (or for high $T$), $T \gg |\mu|$, one finds
	\begin{align}
		n &\simeq \frac{4 \ln 2}{\pi}\, \frac{\mu T}{\hbar^2 v_F^2}\,,
		\label{eqn:n_asym}\\
		n_E &\simeq \frac{6 \zeta(3)}{\pi}\, \frac{T^3}{\hbar^2 v_F^2}\,,
		\label{eqn:n_E_asym}
	\end{align}
	where $\zeta(3) = 1.202$ is the Riemann zeta-function.
	Note that if $n$ is fixed, which is usually the case in the experiment,
	then Eq.~\eqref{eqn:n} implicitly defines
	$\mu$ is a function of $T$. This function $\mu = \mu(n, T)$ can be found by solving  Eq.~\eqref{eqn:n} numerically or (to the leading order) Eq.~\eqref{eqn:n_asym} analytically,
	\begin{equation}
		\mu \simeq \frac{\pi}{4 \ln 2}\, \frac{n}{T}\, \hbar^2 v_F^2\,,
		\label{eqn:mu_asym}
	\end{equation}
	Having obtained $\mu$, one can use Eq.~\eqref{eqn:n_E} to compute the energy density $n_E = n_E(n, T)$ from Eqs.~\eqref{eqn:n_E} or \eqref{eqn:n_E_asym}.
	
	If the electronic temperature $T(t)$ is
	uniform but slowly changing, the linearized equations for the Fourier harmonics of the concentration, energy density, and drift velocity become
	\begin{gather}
		\partial_t n_q + i q n {u}_q =  0\,,
		\label{eqn:continuity_energy_wave}\\
		\partial_t n_{Eq} + \frac{3}{2}\, i q n_{E} {u}_q = 0 \,,
		\label{eqn:heat_energy_wave}\\
		i q v^2_F\left(\frac{n_{E q}}{2} 
		+  e^2 n v_q n_q\right)
		+ \frac32 (n_{E} \partial_t + \partial_t n_{E}) {u}_q = 0\,.
		\label{nonequilibrium_energy_wave}
	\end{gather}
	In these equations we neglected
	the dissipative terms because they are quadratic in the small parameter $q$.
	Similarly, we kept only the leading terms in the adiabatic changing rate $\kappa$. 
	From these equations we can get the third-order differential equation for $n_q$ alone:
	\begin{align}
		&\left( \partial^3_t + b\partial^2_t + c \partial_t \right) n_q(t) =0 \,,
		\label{eqn:n_q_equation}\\
		&b = 2 \partial_t \ln n_{E} \,,
		\:
		c = \frac{2}{3}\, e^2 v^2_F v_q\, \frac{n^2 q^2}{n_{E}} \,.  
	\end{align}
	Therefore, the instantaneous frequency of the energy wave (or ``demon'') is
	$\omega_q = \sqrt{c}$.
	A more accurate expression~\cite{Phan2013} is obtained if the next-order in $q$ terms are retained:
	\begin{align}
		\omega_q = v_F \sqrt{\frac{1}{2}\, q^2
			+ \frac{4\pi}{3} \frac{e^2}{\epsilon}
			\frac{n^2}{n_E}\, q} \,,
		\label{eqn:energy_wave_disperison}
	\end{align}
	which is Eq.~(10) of the main text.
	Formula~\eqref{eqn:energy_wave_disperison} predicts the crossover from $\sqrt{q}$ to linear in $q$ behavior, which was discussed therein.
	Representative plots of the energy wave dispersion
	illustrating its dependence on $T$ are shown in Fig.~\ref{fig:energy_wave_dispersion}.
	
	If $n_E$ is slowly time-dependent, so are the coefficients $b$ and $c$ in Eq.~\eqref{eqn:n_q_equation}.
	This differential equation can be solved within the WKB approximation, which yields Eqs.~(10) and (11) of the main text.
	
	Compared with previous work, our results are in full agreement with those of Ref.~\onlinecite{Phan2013}.
	An equation similar to Eq.~\eqref{eqn:energy_wave_disperison} is Eq.~(43) of Ref.~\onlinecite{Briskot2015}, however the coefficient for the term linear in $q$ differs from that in Eq.~\eqref{eqn:energy_wave_disperison} by a factor of $2\pi$.
	A collective mode with the acoustic dispersion was also discussed in Ref.~\onlinecite{Svintsov2012}. Although the starting equations of that work are mathematically equivalent to our Eqs.~\eqref{eqn:hydro1}--\eqref{eqn:hydro3},
	the final result for the velocity is $0.6 v_F$ instead of $v_F / \sqrt{2}$.
	
	
\end{document}